\documentclass[pra,english,preprintnumbers,amsmath,amssymb,nofootinbib,twocolumn,superscriptaddress]{revtex4-1}
\usepackage[utf8]{inputenc}
\usepackage[T1]{fontenc}
\usepackage{amsmath}
\usepackage{amssymb}
\usepackage{graphicx}
\usepackage{physics}
\usepackage{tabularx}
\usepackage{bm}
\usepackage{ulem}
\usepackage{pifont}

\usepackage{color}
\definecolor{orange}{rgb}{1,0.5,0}

\definecolor{bblue}{rgb}{0.2,0.8,1.0}

\newcommand{\beq}{\begin{equation}}
\newcommand{\eeq}{\end{equation}}
\newcommand{\bea}{\begin{eqnarray}}
\newcommand{\eea}{\end{eqnarray}}

\begin{document}

\title{The virial expansion of attractively interacting Fermi gases \\ 
in 1D, 2D, and 3D, up to fifth order}

\author{Y. Hou}
\affiliation{Department of Physics and Astronomy, University of North Carolina, Chapel Hill, North Carolina 27599, USA}

\author{J. E. Drut}
\affiliation{Department of Physics and Astronomy, University of North Carolina, Chapel Hill, North Carolina 27599, USA}

\date{\today}

\begin{abstract}
The virial expansion characterizes the high-temperature approach to the quantum-classical crossover in any quantum 
many-body system. Here, we calculate the virial coefficients up to the fifth-order of Fermi gases in 1D, 2D, and 3D, with
attractive contact interactions, as relevant for a variety of applications in atomic and nuclear physics.
To that end, we discretize the imaginary-time direction and calculate the relevant
canonical partition functions. In coarse discretizations, we obtain analytic results featuring relationships between 
the interaction-induced changes $\Delta b_3$, $\Delta b_4$, and $\Delta b_5$ as functions of $\Delta b_2$, the latter being exactly 
known in many cases by virtue of the Beth-Uhlenbeck formula. Using automated-algebra methods, we push our 
calculations to progressively finer discretizations and extrapolate to the continuous-time limit.
We find excellent agreement for $\Delta b_3$ with previous calculations in all dimensions and we formulate predictions for 
$\Delta b_4$ and $\Delta b_5$ in 1D and 2D. We also provide, for a range of couplings,
the subspace contributions $\Delta b_{31}$, $\Delta b_{22}$, $\Delta b_{41}$, and $\Delta b_{32}$, which determine the
equation of state and static response of polarized systems at high temperature.
As a performance check, we compare the density equation of state and Tan contact 
with quantum Monte Carlo calculations, diagrammatic approaches, and experimental data where available.
Finally, we apply Pad\'e and Pad\'e-Borel resummation methods to extend the usefulness of the virial coefficients
to approach and in some cases go beyond the unit-fugacity point.
\end{abstract}

\maketitle 

\section{Introduction}

The thermodynamics of interacting fermions at finite density is largely (though not only) 
controlled by the value of the temperature $T$ relative to the Fermi energy scale $\epsilon_F$ or
alternatively the chemical potential $\mu$. For systems with attractive interactions, the regime $\beta \mu \gg 1$,
where $\beta = 1/(k_B T)$, often contains the onset of a superfluid or superconducting transition,
while in the region $\beta \mu \simeq 0$ a crossover regime between quantum and classical physics takes place.
When $z = e^{\beta \mu} \ll 1$, systems are in a dilute, high-temperature regime whose 
thermodynamics is captured by the virial expansion, i.e. an expansion in powers of $z$.
Such an expansion encodes, at a given order $n$, the physics of the $n$-body problem
in the form of virial coefficients. The simplest form of the virial expansion is that of the 
pressure (which is naturally inherited by the density and the compressibility), with corresponding 
coefficients usually denoted by $b_n$.

The applications of the virial expansion in quantum many-body physics have mushroomed in recent years 
with the equally widespread multiplication of ultracold-atom laboratories around the world (see 
e.g.~\cite{Review1, Review2, RevExp, ResonancesReview, ExpReviewLattices}). 
Indeed, as the density is among the easiest thermodynamic observables to determine experimentally~\cite{Exp1,Exp2,Exp3,Exp4}, 
the virial expansion has served as a natural non-perturbative anchor for the results of a variety of theoretical approaches 
to quantum many-body physics (see Ref.~\cite{VirialReview} for a review). Applications of the expansion in 
nuclear physics have also been explored, although to a lesser extent, in the context of finite-temperature neutron 
matter~\cite{SchwenkHorowitz1,SchwenkHorowitz2,SchwenkHorowitz3}.

In the recent work of Ref.~\cite{ShillDrut}, a ``semiclassical'' lattice approximation to the calculation of 
$b_n$ was put forward and applied at leading order for the interaction-induced changes
$\Delta b_3$ and $\Delta b_4$, of spin-$1/2$ Fermi gases in arbitrary spatial dimensions. 
The results were compared with quantum Monte Carlo (QMC) calculations in 1D and 2D as well as diagrammatic approaches in 2D.
Notably, while that approximation corresponds to a maximally coarse Trotter-Suzuki factorization of the imaginary-time 
evolution operator, the answers were in good quantitative agreement.
Reference~\cite{HouEtAl} carried out the approximation one order further and up to $\Delta b_7$, with similar results when 
comparing with previous calculations (where available). Tests of the coarse approximation for $\Delta b_3$ and $\Delta b_4$ in a harmonic
trapping potential were also obtained, in Ref.~\cite{MorrellBergerDrut}, which showed remarkable qualitative and semi-quantitative
agreement with prior calculations as a function of the trap frequency.

Encouraged by such a positive experience, here we take the calculation up to $\Delta b_5$ by refining 
the factorization as much as possible for each virial coefficient. We extend our work to 1D and 2D, focus still on 
attractive contact interactions, and present applications to the equation of state and Tan contact.
For completeness, we also discuss aspects of the 3D system here that we did not discuss in our previous work~\cite{HouDrut}, namely
the isothermal compressibility. 
In all cases, we present the decomposition of $\Delta b_n$ in terms of its subspace contributions,
namely $\Delta b_{31}$ and $\Delta b_{22}$ for $\Delta b_4$, and $\Delta b_{41}$ and $\Delta b_{32}$ for $\Delta b_5$.
These subspace contributions are not often discussed, but they are important: they determine the thermodynamics (energy, entropy, 
density, static response, and so on) of the polarized systems. Furthermore, as we will see, $\Delta b_{mj}$ displays simpler and more 
systematic behavior than $\Delta b_n$ as a function of the order in the expansion and the coupling strength.. Finally, we take a step further 
and analyze the virial coefficients using resummation techniques, namely the Pad\'e and Borel-Pad\'e methods, which substantially extend the 
applicability (in a practical sense) of the virial expansion in all the cases we studied.

The remainder of this paper is organized as follows. Section~\ref{Sec:HVE} presents the Hamiltonian we focus on
along with the formal elements of the virial expansion, and explains the basics of our approach, which is
based on the discretization of imaginary time by a Trotter-Suzuki factorization of the Boltzmann weight. 
Section~\ref{Sec:AnalyticResults} presents our analytic formulae for canonical partition functions in Sec.~\ref{Sec:ResultsA} and virial coefficients for arbitrary spatial dimension in Sec.\ref{Sec:ResultsB}. 
In Sec.~\ref{Sec:ExtrapolatedResults}, we show our results from extrapolating to the continuous-time limit. Specifically,
Secs.~\ref{Sec:ResultsC} through \ref{Sec:ResultsH} show our results for virial coefficients and corresponding applications in 
1D, 2D, and 3D. As a way to extend the application of the virial expansion, we use Pad\'e and Pad\'e-Borel resummation
techniques, which we comment on in Sec.~\ref{Sec:ResumTech}. Finally, we summarize and conclude in Sec.~\ref{Sec:Conclusion}.

\section{\label{Sec:HVE}Hamiltonian, virial expansion, and method}

The simplest interacting effective theory one can write
for nonrelativistic spin-$1/2$ fermions in $d$ spatial dimensions has
a Hamiltonian given by $\hat H = \hat T + \hat V$,
where
\bea
\label{Eq:T}
\hat T \!=\! \sum_{s=\uparrow,\downarrow}{\int{d^d x\,\hat{\psi}^{\dagger}_{s}({\bf x})\left(-\frac{\hbar^2\nabla^2}{2m}\right)\hat{\psi}_{s}({\bf x})}},
\eea
and
\bea
\label{Eq:V}
\hat V \!=\! - g_{d}\! \int{d^d x\,\hat{n}_{\uparrow}({\bf x})\hat{n}_{\downarrow}({\bf x})},
\eea
where the field operators $\hat{\psi}_{s}, \hat{\psi}^{\dagger}_{s}$ are fermionic fields for particles of spin $s=\uparrow,\downarrow$, 
and $\hat{n}_{s}({\bf x})$ are the coordinate-space densities. In the remainder of this work, we will take $\hbar = k_\text{B} = m = 1$. 
To regularize the interaction, we will put the Hamiltonian on a spatial lattice of spacing $\ell$, whose calibration will be 
determined by our renormalization condition, as explained below.

As mentioned above, the virial expansion is an expansion around the dilute limit
$z\to 0$, where $z=e^{\beta \mu}$ is the fugacity, $\beta$ is the inverse temperature, 
and $\mu$ the chemical potential coupled to the total particle number operator $\hat N$.
The coefficient accompanying the $n$-th power of $z$ in the expansion of the grand-canonical 
potential $\Omega$ is the virial coefficient $b_n$:
\beq
-\beta \Omega = \ln {\mathcal Z} = Q_1 \sum_{n=1}^{\infty} b_n z^n,
\eeq
where
\beq
\mathcal Z = \tr \left[e^{-\beta (\hat H - \mu \hat N)}\right] = \sum_{N=0}^{\infty} z^N Q_N,
\eeq 
is the grand-canonical partition function. $Q_N$ is the $N$-body partition function, $b_1 = 1$, and the higher-order 
coefficients are given by
\bea
Q_1 b_2 &=& Q_2 - \frac{Q_1^2}{2!},\\
Q_1 b_3 &=& Q_3 - b_2 Q_1^2  - \frac{Q_1^3}{3!},\\
Q_1 b_4 &=& Q_4 -  \left(b_3 + \frac{b_2^2}{2}\right) Q_1^2 -b_2\frac{Q_1^3}{2!}  - \frac{Q_1^4}{4!}, \\
Q_1 b_5 &=& Q_5 - (b_4  + b_2 b_3 ) Q_1^2 - \left (b_2^2  + b_3 \right )\frac{Q_1^3}{2} \nonumber \\
&&- b_2 \frac{Q_1^4}{3!}  - \frac{Q_1^5}{5!},
\eea
etcetera. The noninteracting virial coefficients for nonrelativistic fermions in $d$ spatial dimensions are $b^{(0)}_n = (-1)^{n+1} n^{-(d+2)/2}$.

The highest power of $Q_1$ does not involve any virial coefficients and therefore always disappears in the interaction change $\Delta b_n$:
\bea
\label{Eq:DeltabnDeltaQn}
Q_1 \Delta b_2 &=& \Delta Q_2 \\
Q_1 \Delta b_3 &=& \Delta Q_3 - Q_1^2 \Delta b_2,\\
Q_1 \Delta b_4 &=& \Delta Q_4 -  \Delta\left(b_3 + \frac{b_2^2}{2}\right) Q_1^2 -\frac{\Delta b_2}{2} Q_1^3, \\
Q_1 \Delta b_5 &=&\Delta  Q_5 - \Delta (b_4  + b_2 b_3 ) Q_1^2 \nonumber \\ 
&&- \frac{1}{2}\Delta \left (b_2^2  + b_3 \right )Q_1^3 - \frac{\Delta b_2}{3!} Q_1^4.
\eea

Furthermore, in terms of the partition functions $Q_{MN}$ of $M$ particles of one type and $N$ of the other type, we have
\bea
\Delta Q_2 &=& \Delta Q_{11}, \\
\Delta Q_3 &=& 2\Delta Q_{21}, \\
\Delta Q_4 &=& 2\Delta Q_{31} + \Delta Q_{22}, \\
\Delta Q_5 &=& 2\Delta Q_{32} + 2\Delta Q_{41}.
\label{Eq:DeltaQnDeltaQnm}
\eea
Therefore, the main complexity in the calculations presented below is in computing
the few $\Delta Q_{MN}$ shown above within the Trotter-Suzuki factorization of the imaginary-time
evolution operator, which we discuss next.

\subsection{Trotter-Suzuki factorization}

To carry out our calculations, we factorize the imaginary time evolution operator in the style of Trotter-Suzuki, 
such that 
\beq
\label{Eq:TSFactorization}
e^{-\beta (\hat T + \hat V)} \simeq \left( e^{-\beta \hat T / k}e^{-\beta \hat V/ k} \right)^k,
\eeq
where setting $k=1$ defines the coarsest possible discretization.
In this work we will take $k \gg 1$ and as far as possible. For low $k$, however,
we carry out analytic calculations leading to explicit formulas which are useful
for cross-checks across dimensions, as they feature the spatial dimension $d$ as a parameter.

For $k=1$ we have
\beq
\label{Eq:LOSCLA}
e^{-\beta (\hat T + \hat V)} \simeq e^{-\beta \hat T }e^{-\beta \hat V},
\eeq
which is equivalent to neglecting $[\hat T , \hat V]$ and higher-order commutators; for that reason we 
have called this a semiclassical expansion in previous works. 
In all cases of interest here, the expressions will appear inside a trace, such that the error is pushed to 
$\mathcal O(1/k^2)$. Indeed, as far as the trace is concerned, Eq.~(\ref{Eq:LOSCLA}) is equivalent
to the more accurate, symmetric decomposition
\beq
e^{-\beta (\hat T + \hat V)} \simeq \left(e^{-\beta \hat T/(2k) }e^{-\beta \hat V/k} e^{-\beta \hat T/(2k)} \right)^k,
\eeq
whose error is $\mathcal O(1/k^2)$.

\section{\label{Sec:AnalyticResults}Analytic results}

In this section we present our main analytic results. In Sec.~\ref{Sec:ResultsA} we present partition functions for $k=2$ 
and in Sec.~\ref{Sec:ResultsB} virial coefficients up to the fifth order for $k=1$ and $k=2$. In all cases the results
feature the spatial dimension $d$ as a variable.
Our analytic results are shown as examples and cross-checks for the method, and for future reference.

\subsection{\label{Sec:ResultsA}Canonical partition functions for $k=2$}

As the simplest nontrivial example, we work out in detail $Q_{11}$ which determines $b_2$ and therefore
plays a central role in our method. For $k=2$, the calculation
begins as follows:
\bea
Q_{11} &=& \sum_{{\bf p}_1 {\bf p}_2} \langle {\bf p}_1 {\bf p}_2 |   e^{ -\frac{\beta \hat T}{2}}e^{-\frac{\beta \hat V}{2}}e^{-\frac{\beta \hat T}{2}}e^{-\frac{\beta \hat V}{2}}  | {\bf p}_1 {\bf p}_2 \rangle \\
&&\!\!\!\!\!\!\!\!\!\!\!\! = \sum_{{\bf p}_1 {\bf p}_2{\bf p}_3 {\bf p}_4} e^{- \beta ({\bf p}_1^2 + {\bf p}_2^2)/4 m } e^{- \beta ({\bf p}_3^2 + {\bf p}_4^2)/4m } \times \nonumber\\
&&  \mel{{\bf p}_1 {\bf p}_2}{e^{-\beta \hat V/2}}{{\bf p}_3 {\bf p}_4} \mel{{\bf p}_3 {\bf p}_4}{e^{-\beta \hat V/2}}{{\bf p}_1 {\bf p}_2}. \nonumber
\eea
The next step is to insert a coordinate-space completeness relation and use the following identity:
\bea
e^{-\beta \hat V/2} | {\bf x}_1 {\bf x}_2 \rangle &=& \prod_{\bf z} (1 + C\hat{n}_{\uparrow}({\bf z})\hat{n}_{\downarrow}({\bf z}))| {\bf x}_1 {\bf x}_2 \rangle  \\
&&\!\!\!\!\!\!\!\!\!\!\!\! = | {\bf x}_1 {\bf x}_2 \rangle + C \sum_{\bf z}\delta_{{\bf x}_1,{\bf z}} \delta_{{\bf x}_2, {\bf z}}| {\bf x}_1 {\bf x}_2 \rangle \nonumber \\
&&\!\!\!\!\!\!\!\!\!\!\!\! = \left[1 + C \delta_{{\bf x}_1,{\bf x}_2} \right]| {\bf x}_1 {\bf x}_2 \rangle \nonumber,
\eea
where $C = (e^{\beta g_{d}/2} - 1)\ell^{d}$ and we used the fermionic relation $\hat{n}^2_s = \hat{n}_s$. Note that the series in 
powers of $C$ terminates at linear order for this particular state in which there is only one particle for one of the species (regardless
of how many particles of the other species are present).
The $C$-independent term yields 
the noninteracting result, such that
\bea
\!\!\!\!\Delta Q_{11} &=& \sum_{{\bf P} {\bf P}^\prime {\bf X} {\bf X}^\prime} e^{- \beta ({\bf P}^2 + {\bf P^\prime}^2) / 4m} \times \\
&& \braket{{\bf P}}{{\bf X}} \braket{{\bf X}}{{\bf P^\prime}} \braket{{\bf P^\prime}}{{\bf X^\prime}} \braket{{\bf X^\prime}}{{\bf P}} \times \nonumber \\
&& \left [ C \left (\delta_{{\bf x}_1, {\bf x}_2} + \delta_{{\bf x^\prime}_1, {\bf x^\prime}_2} \right )
 + C^2 \left (\delta_{{\bf x}_1, {\bf x}_2} \delta_{{\bf x^\prime}_1, {\bf x^\prime}_2} \right ) \right ] \nonumber
\eea
where \( {\bf P} \) is the shorthand for the set of momentum variables \( {\bf p}_1 \) and \( {\bf p}_2 \) and \( {\bf P}^2 = {\bf p}_1^2 + {\bf p}_2^2 \), and
similarly for \( {\bf P}' \), \( {\bf X} \), \( {\bf X}' \).

Using a plane wave basis, $|\langle {\bf x}_1 {\bf x}_2 | {\bf p}_1 {\bf p}_2 \rangle |^2 = 1/V^2$,
where $V = L^d$ in $d$ spatial dimensions and $L$ is the linear extent of the system, and we then find
\bea
\Delta Q_{11} &= & C \frac{f_1}{V} + C^2 \frac{f_2}{V^2},
\eea
where $f_1 = 2 Q_{10}^2$, with
\beq
Q_{10} = \sum_{{\bf p}_1} e^{-\beta p_1^2/2m},
\eeq
and
\beq
f_2 = \sum_{{\bf p}_1 {\bf p}_2 {\bf p}_3} e^{-\beta \left[ {\bf p}_1^2 + {\bf p}_2^2 + {\bf p}_3^2 + ({\bf p}_1 + {\bf p}_2 - {\bf p}_3)^2 \right] / 4m}.
\eeq

In the continuum limit, in $d$ spatial dimensions,
\beq
Q_{10} = \left(\frac{L}{\lambda_T}\right)^d,
\eeq
\beq
f_1 = 2 \left(\frac{L}{\lambda_T}\right)^{2d}, \ \ \ \ \ f_2 = \left(\frac{L}{\lambda_T}\right)^{3d} 2^\frac{d}{2}.
\eeq
where $\lambda_T= \sqrt{2\pi\beta}$ is the thermal wavelength. 

Thus, 
\beq
\Delta Q_{11} = 2 \left(\frac{L}{\lambda_T}\right)^d \left[ \frac{C}{\lambda_T^d} +  2^{\frac{d}{2}-1} \left(\frac{C}{\lambda_T^d}\right)^2 \right ].
\eeq
such that
\beq
\Delta b_2 = \frac{\Delta Q_{11}}{Q_{1}} = \frac{C}{\lambda_T^d} + 2^{\frac{d}{2}-1}\left(\frac{C}{\lambda_T^d}\right)^2,
\eeq
where we have used $Q_1 = 2 Q_{10}$. 

As explained in the introduction, only a few canonical partition functions enter
in $\Delta b_n$. For $n = 3,4, 5$ all we need is
$\Delta Q_{21}$, $\Delta Q_{31}$, $\Delta Q_{22}$, $\Delta Q_{41}$, $\Delta Q_{32}$. Of these,
the last two are the most mathematically demanding. Below we present a sample of our analytic results for 
$k=2$ for selected partition functions (excluding $\Delta Q_{41}$, $\Delta Q_{32}$) in the continuum limit.

Calculating $\Delta b_3$ requires $\Delta b_2$ and the following result:
\bea
\frac{2\Delta Q_{21}}{Q_1} &=& \frac{C}{\lambda_T^d}
    \left[ - 2^{1- \frac{d}{2}}  + \left( \frac{L}{\lambda_T} \right)^{d}  \right]
    \nonumber \\
    &&\!\!\!\!\!\!+ \left(\frac{C}{\lambda_T^d}\right)^2 \left[
      \left(1 - \frac{2^{d+1}}{5^\frac{d}{2}}\right)
      + 2^\frac{d}{2}\left(\frac{L}{\lambda_T}\right)^{d}
  \right].
\eea
Both of the contributions displaying explicit dependence on $L/\lambda_T$ will
cancel out in the final expression for $\Delta b_3$, giving a volume-independent
result.
\begin{widetext}

Calculating $\Delta b_4$ requires $\Delta b_2$, $\Delta b_3$, and the following two results:
\bea
\frac{2\Delta Q_{31}}{Q_1} &=& \frac{C}{\lambda_T^d} 
\left[ 
\frac{2}{3^\frac{d}{2}}
- \frac{3}{2^\frac{d}{2}} \left( \frac{L}{\lambda_T} \right)^{d} 
+ \left( \frac{L}{\lambda_T} \right)^{2d}  \right]
+
\left(\frac{C}{\lambda_T^d}\right)^2 
\left[ 
    \frac{1}{3^\frac{d}{2}}
  + \left( \frac{1}{2} -  \frac{2^{d+1}}{5^{\frac{d}{2}}} \right) \left( \frac{L}{\lambda_T} \right)^{d} 
  + \frac{1}{2^{1-\frac{d}{2}}} \left( \frac{L}{\lambda_T} \right)^{2d} 
\right] ,
\eea

\bea
\frac{\Delta Q_{22}}{Q_1} &=& \frac{C}{\lambda_T^d}
\left[
  2^{-d}
- 2^{1-\frac{d}{2}} \left( \frac{L}{\lambda_T} \right)^{d} 
+ \left( \frac{L}{\lambda_T} \right)^{2d} \right] \nonumber \\
&+& \left(\frac{C}{\lambda_T^d}\right)^2
\left[
   \left( 2^{-d-1} + \frac{2}{3^\frac{d}{2}} - \frac{3}{2^\frac{d}{2}} \right)
  + 2 \left( 1 - \frac{2^{d}}{5^\frac{d}{2}}  \right) \left( \frac{L}{\lambda_T} \right)^{d} 
  + 2^{\frac{d}{2}-1} \left( \frac{L}{\lambda_T} \right)^{2d} 
\right] \nonumber \\
&+& \left(\frac{C}{\lambda_T^d}\right)^3
\left[ 
   \left( 1 + 2^{1-\frac{d}{2}} - \frac{2^{2+d}}{5^\frac{d}{2}} \right)
   + 2^\frac{d}{2}\left( \frac{L}{\lambda_T} \right)^{d} 
\right]
+ \left(\frac{C}{\lambda_T^d}\right)^4 
\left[ 
   \left(\frac{3}{4} - \frac{2^{d}}{3^\frac{d}{2}} \right)
   + 2^{-2+d}\left( \frac{L}{\lambda_T} \right)^{d} 
\right]. 
\eea
\end{widetext}
As mentioned above, also in this case, the explicit dependence on $L/\lambda_T$
will be cancelled in the final expression for $\Delta b_4$. Only the volume-independent terms will
remain.

In the next section, we use the above expressions to assemble the calculation of 
$\Delta b_3$ and $\Delta b_4$, using Eqs.~(\ref{Eq:DeltabnDeltaQn}) through~(\ref{Eq:DeltaQnDeltaQnm})
at $k=2$. We will, in fact, go beyond the above expressions and present results for $\Delta b_5$ as well,
and extend the whole analysis to $k=3,4$ and extrapolate to the continuous-time limit.
As the equations for $k \geq 3$ are much too long to be displayed in the present format in a useful manner,
we have written a Python code that evaluates our formulas for $k=1,2,3,4$, available as part of our Supplemental Material~\cite{SupMat}.
[Note however that, beyond the analytic results presented here, where the spatial dimension $d$ appears explicitly,
we will not show continuous cross-dimensional results for the virial coefficients in this work but rather focus on 1D, 2D, and 3D. 
The Supplemental Materials contain the numerical values of the virial coefficients in 1D, 2D and 3D, after extrapolation to 
the large-$k$ limit.]

\subsection{\label{Sec:ResultsB}Virial coefficients: Analytic results at $k=1$ and $k=2$ across dimensions}
Previous work~\cite{ShillDrut, HouEtAl} calculated the virial coefficients at $k=1$,
which yields, for a fermionic two-species system with a contact interaction, in $d$ spatial dimensions,
\bea
\label{Eq:Db2ToDb4LOSCLA_Db2}
\Delta b_3 &=& -2^{1-\frac{d}{2}} \Delta b_2, \\
\Delta b_4 &=& 2(3^{-\frac{d}{2}}\!\!+\! 2^{-d-1} ) \Delta b_2 \nonumber \\
&&+ 2^{1-\frac{d}{2}}(2^{-\frac{d}{2}-1}\!\!-\!\! 1) (\Delta b_2)^2,
\eea
where we have corrected the coefficient of $(\Delta b_2)^2$ relative to Ref.~\cite{ShillDrut}.
Also at $k=1$, but going beyond the work of Ref.~\cite{ShillDrut}, Ref.~\cite{HouEtAl} found
\bea
\Delta b_5 &=& -(2^{-d} + 6^{-\frac{d}{2}}) \Delta b_2 \nonumber \\
&&+ 4(2^{-d} + 3^{-\frac{d}{2}} - 7^{-\frac{d}{2}}) (\Delta b_2)^2,
\eea
which we show here for completeness as we will calculate $\Delta b_5$ at higher $k$.

As part of our main results, we have extended the above calculations to higher $k$ for 
$\Delta b_3$, $\Delta b_4$, and $\Delta b_5$. For $k=2$ one can write down explicit formulas that easily fit on a sheet of paper:
\bea
\label{Eq:Db2ToDb4NLOSCLA}
\Delta b_2 &=& \tilde C + 2^{\frac{d}{2}-1} \tilde C^2, \\
\Delta b_3 &=& -2^{1-\frac{d}{2}} \tilde C + \left( 1 - \frac{2^{1+d}}{5^\frac{d}{2}} \right) \tilde C^2, \\
\Delta b_4 &=& 2(3^{-\frac{d}{2}} + 2^{-d-1} ) \tilde C + \left( 3^{1-\frac{d}{2}} + 2^{-d-1} \!\!-\!\! \frac{3}{2^\frac{d}{2}} \right) \tilde C^2 \nonumber\\
&+& \left(1 + 2^{1-\frac{d}{2}} - \frac{2^{d+2}}{5^\frac{d}{2}} \right) \tilde C^3 + \left( \frac{3}{4} - \frac{2^d}{3^\frac{d}{2}} \right) \tilde C^4, \\
\Delta b_5 &=& -\left(2^{1-d} + \frac{2^{1-\frac{d}{2}}}{3^\frac{d}{2}}\right) \tilde C \nonumber \\
&+& \left(\frac{7}{2^d} - \frac{2^{1+d}}{3^{\frac{3d}{2}}} + \frac{7}{3^\frac{d}{2}} - \frac{2}{7^\frac{d}{2}} - \frac{2^{1+d}}{11^\frac{d}{2}}-3\cdot\frac{2^{1+d}}{19^\frac{d}{2}} \right) \tilde C^2 \nonumber \\
&+& \left[2^{1-d} - 2^{1-\frac{d}{2}} + 4 \cdot 3^{1-\frac{d}{2}} \right. \nonumber \\
&&  \left. - 2^{\frac{d}{2}+2} \left( \frac{2}{3^{d}} + 5^{-\frac{d}{2}} - 7^{-\frac{d}{2}} \right) \right]  \tilde C^3 \nonumber \\
&+& \left( 1 + 2^{2 - \frac{d}{2}} - 2^{1 + d}\, 3^{1 - d} - 3\cdot \frac{2^{1 + d}}{5^\frac{d}{2}} \right. \nonumber \\
&& \left.- \frac{4^d}{3^{d}\, 5^{\frac{d}{2}}}  + \frac{3^{1 - \frac{d}{2}} 4^d}{7^{\frac{d}{2}}} + 3 \cdot \frac{2^{1 + 2 d}}{29^\frac{d}{2}} \right)  \tilde C^4,
\eea
where $\tilde C= (e^{\beta g_{d}/2} - 1)\ell^d/{\lambda_T^d}$.
Although we used these $k=2$ results in a previous work~\cite{HouEtAl}, we did not include the above formulas explicitly.

Note that, while the above expressions resemble truncated power series in $\tilde C$, they actually 
display the full answer for $k=2$.
Furthermore, we note that as in the $k=1$ case, in the $k=2$ case $\Delta b_2$ is always positive and 
$\Delta b_3$ is always negative, for positive $\tilde C$ in $d = 1,2,3$. The behavior
of $\Delta b_4$ and $\Delta b_5$, however, is less obvious, which is at least in part because
they are the result of competing subspace contributions (see Ref.~\cite{HouDrut} and our discussion below).
\begin{figure}[t]
  \begin{center}
   \includegraphics[scale=0.57]{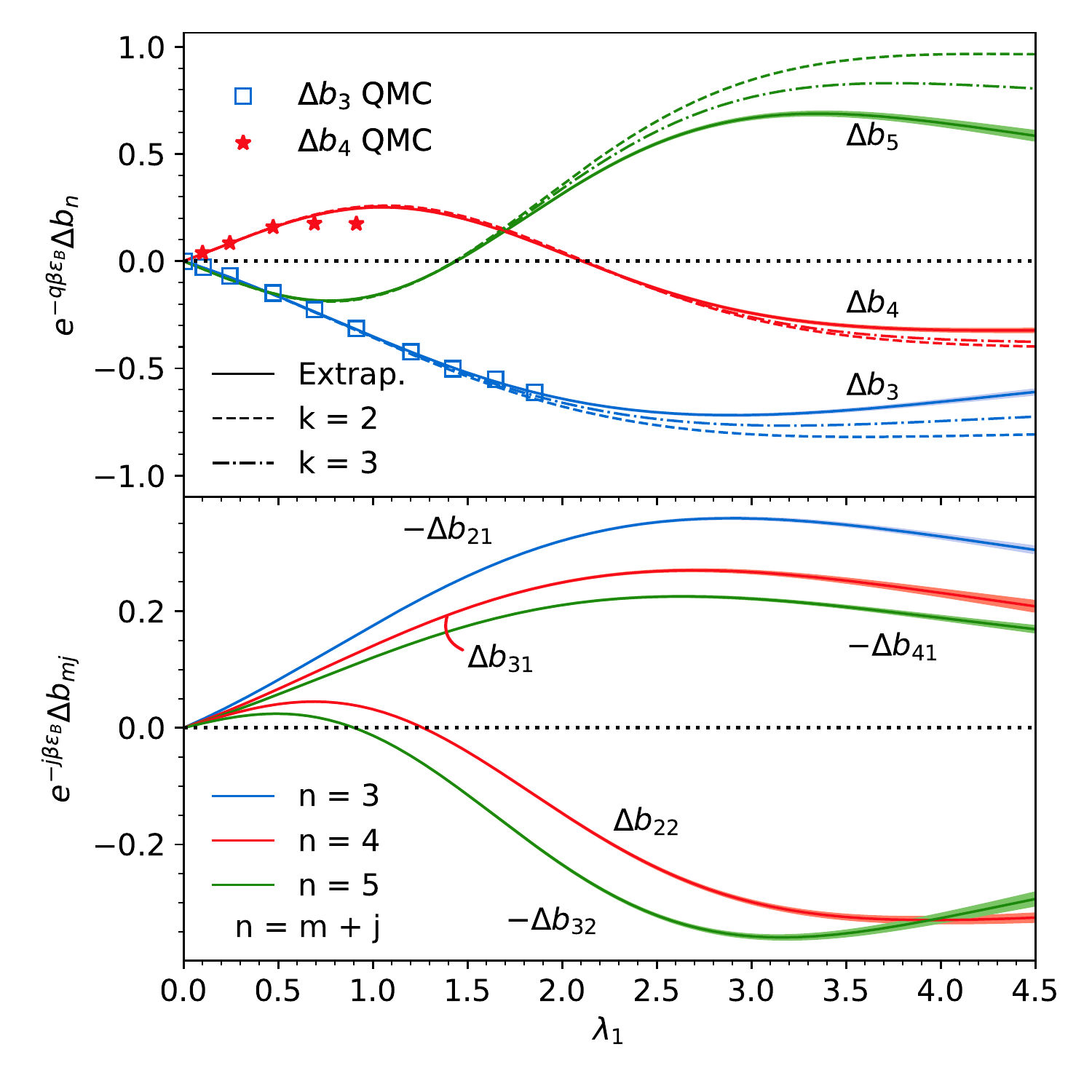}
  \end{center}
  \caption{{\bf Top}: Virial coefficients $\Delta b_n$ for $n=2-5$ for the 1D attractive Fermi gas as a function of the dimensionless coupling $\lambda_1$.
    To fit the scale, the \( \Delta b_n \)s are scaled by \( \exp(-q \beta \epsilon_b)  \), where \( q \) is the maximum number of spin-$\uparrow\downarrow$ pairs.
    The $k=2$ results are shown as dashed lines, $k=3$ as dashed-dotted lines, and the $k\to\infty$ extrapolation with solid lines. 
Results for $\Delta b_3$ appear in blue, $\Delta b_4$ in red, and $\Delta b_5$ in green.
Blue squares and red stars show the QMC results for $\Delta b_3$ and $\Delta b_4$, respectively, from Ref.~\cite{ShillDrut}.
{\bf Bottom}: Subspace contributions \( \Delta b_{mj} \) as functions of the coupling strength,
  scaled by \( \exp(-j \beta \epsilon_b) \).
 Our results are shown as error bands, color-coded as in the top plot by $n = m+j$: 
 blue for $-\Delta b_{21}$, red for $\Delta b_{31}$ and $\Delta b_{22}$, and green for 
 $-\Delta b_{41}$ and $-\Delta b_{32}$.
 Specific cases are inverted in sign for clarity and to avoid overlaps.
   }
  \label{Fig:1Dbn}
\end{figure}
Solving for $\tilde C$ in terms of $\Delta b_2$ at $k=2$, we find
\beq
\label{Eq:Ctilde}
\tilde C = 2^{-\frac{d}{2}}\left(\sqrt{1 + 2^{\frac{d}{2} + 1} \Delta b_2} - 1 \right),
\eeq
where we have chosen the solution that yields a real and positive value for $\tilde C$, which 
corresponds to attractive interactions and thus positive $\Delta b_2$.
Using that result yields $\Delta b_3$, $\Delta b_4$, and $\Delta b_5$ in terms of $\Delta b_2$.
In the sections that follow we will use the above formulas and their generalizations to $k=3$ and beyond
to display the behavior of virial coefficients as $k$ is increased and compare those results with
the extrapolations to the large-$k$ limit.

\section{\label{Sec:ExtrapolatedResults}Extrapolated Results}

In this section we focus on the physically relevant cases of 1D, 2D, and 3D attractive Fermi gases, 
extrapolating to the large-$k$ (i.e. continuous-time) limit. Specifically, we calculated the interaction-induced changes in the virial 
coefficients \( \Delta b_3 \) up to \( k = 21 \), \( \Delta b_4 \) up to \( k = 12 \) and \( \Delta b_5 \) up to \( k=9 \), and based on those 
results we extrapolated to \( k \to \infty \) using the techniques discussed in Ref.~\cite{HouDrut}. 

In each spatial dimension, we discuss applications centered around measurable quantities
such as the density equation of state, Tan contact, and the isothermal compressibility. The applicability and usefulness
of the calculated coefficients extends well beyond those observables, however, as the $\Delta b_n$ also determine the high-temperature 
behavior of a broad suite of thermodynamic quantities such as the energy, entropy, pressure, magnetic susceptibility, etc. Furthermore, we 
focus largely on unpolarized systems (the exceptions being the density equation of state in 1D and 3D at unitarity), but provide the subspace 
decomposition of \( \Delta b_4 \) and \( \Delta b_5 \), which extend the applicability of our results to the polarized-system version of the 
aforementioned thermodynamic quantities.

\subsection{\label{Sec:ResultsC}Virial coefficients in 1D}

To calculate the virial coefficients of attractively interacting fermions in 1D we used 
the technique described above combined with the exact Beth-Uhlenbeck result~\cite{BU, EoS1D}
\beq
\Delta b_2^\text{1D,exact} = 
-\frac{1}{2\sqrt{2}} + \frac{e^{\lambda_1^2/4}}{2\sqrt{2}}[1 + \text{erf}(\lambda_1/2)].
\eeq
This equation, together with Eq.~(\ref{Eq:Ctilde}) (and its counterparts at higher $k$), 
allows us to obtain $\tilde C$ as a function of the dimensionless physical coupling 
$\lambda_1 = 2\sqrt{\beta}/a_0$, where $a_0$ is the 1D scattering length.

The results of our 1D calculations are shown in Fig.~\ref{Fig:1Dbn} and are in excellent agreement with the QMC data 
of Ref.~\cite{ShillDrut} for $\Delta b_3$ and $\Delta b_4$. For the latter, the QMC calculations are very limited and stop beyond 
at $\lambda_1 \simeq 1.0$ (where the QMC method begins to break down), whereas our present results go well beyond that region. 
Because a two-body bound state forms in this system as soon as the interaction is turned on,
the virial coefficients will tend to grow exponentially with the binding energy (as is evident at second order from the 
Beth-Uhlenbeck formula). To capture that behavior, 
we scaled our virial coefficients by the inverse Boltzmann weight
$\exp(-q \beta \epsilon_B)$ of the available particle pairs $q$ (i.e. $q=1$ for \( \Delta b_3 \) and $q=2$ for \( \Delta b_4 \) and  \( \Delta b_5 \)),
where $\epsilon_B = 1/a_0^2$ is the binding energy of the two-body system.
The resulting mild behavior at strong coupling shows that indeed the scaling factor captures the shape of  
\( \Delta b_n \) as the coupling is increased. Beyond that leading contribution, however, \( \Delta b_3 \) is controlled
by the atom-dimer scattering properties, just as \( \Delta b_4 \) is controlled by dimer-dimer properties, and so on.
In all the dimensions explored here, we found that $\Delta b_3$ is of constant sign, 
whereas $\Delta b_4$ and $\Delta b_5$ change sign at strong enough coupling, as a result of the
competition between positive and negative contributions coming from the fixed-polarization subspaces $\Delta b_{mj}$.

\begin{figure}[t]
  \begin{center}
   \includegraphics[scale=0.53]{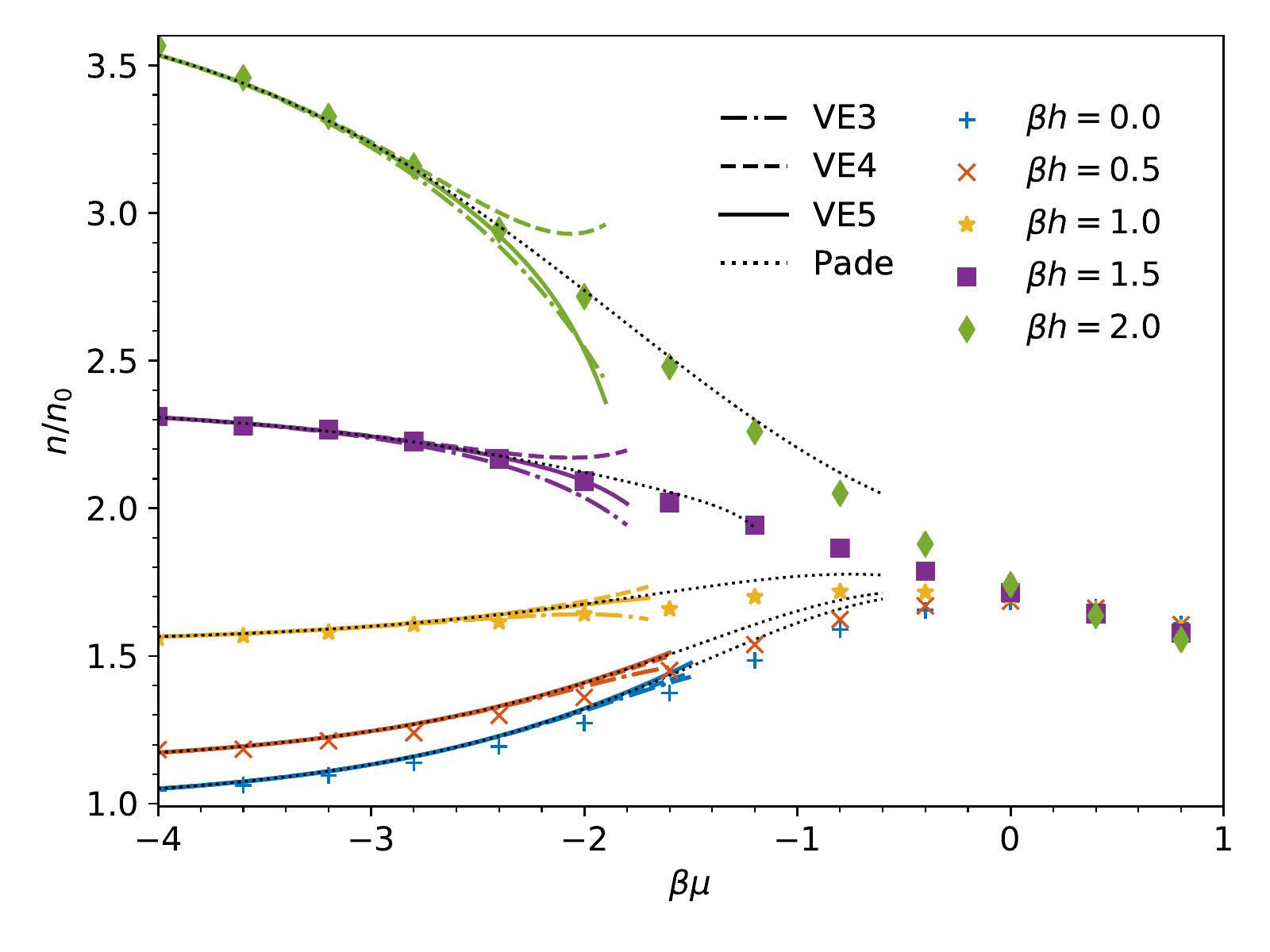}
  \end{center}
  \caption{Density equation of state \( n \) of the attractive 1D Fermi gas, shown in units of the noninteracting, unpolarized counterpart \( n_0 \) with coupling strength \( \lambda_1 = 1.0 \), for several values of the chemical potential difference $\beta h = \beta (\mu_\uparrow - \mu_\downarrow)/2$.
    The colored symbols are complex-Langevin results from Ref.~\cite{LoheacDrutBraun} and the colored lines show the virial expansion at 
    various orders: dashed-dotted line at third order (VE3), dashed line at fourth order (VE4), and solid line at fifth order (VE5).
    The black dotted line is the result of the $[3/2]$ Pad\'e resummation (see Sec.~\ref{Sec:ResumTech}).
    The Pad\'e approximant for \( \beta h = 1.5 \) contains a pole close to \( \beta \mu = -1.2 \); the corresponding dotted line is therefore cut off at that value.
    The limiting value for $\beta \mu \to -\infty$ is known exactly and is given by $\cosh(\beta h)$.
  }
  \label{Fig:density-polarized-1d}
\end{figure}

The $\Delta b_{mj}$ are shown in the bottom panel of Fig.~\ref{Fig:1Dbn}. There is naturally only one subspace contributing 
at third order: $\Delta b_3 = 2\Delta b_{21}$. However, $\Delta b_4 = \Delta b_{22} + 2 \Delta b_{31}$ and 
$\Delta b_5 = 2\Delta b_{32} + 2 \Delta b_{41}$. Furthermore, in each of the latter, the subspace terms enter with similar magnitudes but 
opposing signs, and thus compete to determine the sign and size of $\Delta b_n$.
It is hard to miss that $|\Delta b_{m1}|$ follows essentially the same trend as a function of the coupling across all $m$,
and that trend becomes increasingly suppressed for increasing $m$. A similar observation applies to $|\Delta b_{m2}|$.
The suppression with increased polarization is easy to understand: at large $m$ both $|\Delta b_{m1}|$ and $|\Delta b_{m2}|$
must approach the noninteracting values (i.e. zero) as the majority of the particles do not interact due to Pauli blocking.

\subsection{\label{Sec:ResultsD}Applications in 1D}

\subsubsection{Density equation of state at finite polarization}

In the virial expansion, the density equation of state of a polarized system becomes a double series expansion 
in powers of the fugacity of each spin $z_s = \exp(\beta \mu_s)$, $s = \uparrow, \downarrow$. More specifically, the grand 
thermodynamic potential $\Omega = \Omega_0 + \Delta \Omega$, relative to its noninteracting counterpart $\Omega_0$
becomes
\begin{equation}
-\beta \Delta \Omega = Q_1 \sum_{n=2}^{\infty} \sum_{\substack{m,j > 0 \\ m + j  = n}} \Delta b_{mj} z_{\uparrow}^m z_{\downarrow}^j,
\end{equation}
such that the particle number density for spin-$\uparrow$ is given by
\beq
\label{Eq:DensityUp}
\lambda_T^d n_\uparrow = \lambda_T^d n_{\uparrow,0} + 2 \sum_{n=2}^{\infty} \sum_{\substack{m,j > 0 \\ m + j  = n}} m \Delta b_{mj} z_{\uparrow}^m z_{\downarrow}^j,
\eeq
where $\lambda_T= \sqrt{2\pi\beta}$ is the thermal wavelength, $\lambda_T^d n_{\uparrow,0} = f_{d/2}(z_{\uparrow})$ is the noninteracting value in $d$ 
dimensions, \( f_{d/2}(z) = \text{Li}_{d/2}(-z) \) is the Fermi-Dirac function, and 
$\text{Li}_{s}(x)$ is the polylogarithm function. An analogous expression to Eq.~(\ref{Eq:DensityUp}) holds for $\lambda_T^d n_{\downarrow}$.

In Fig.~\ref{Fig:density-polarized-1d} we show our estimates for the density equation of state at finite polarization in the virial expansion,
for attractively interacting fermions in 1D.
We compare our results with those of Ref.~\cite{LoheacDrutBraun} obtained with the complex Langevin method.
The fifth-order virial expansion provides a modest improvement over the third and fourth orders. However, for all the available
polarizations the agreement with the data is reasonable as long as $\beta \mu$ is sufficiently small. Resummation techniques (see below)
like Pad\'e approximants extend that agreement (which while not perfect, it is better than qualitative) with the available data over a wider $\beta \mu$ region.

\subsubsection{Contact}
\begin{figure}[t]
  \begin{center}
   \includegraphics[scale=0.53]{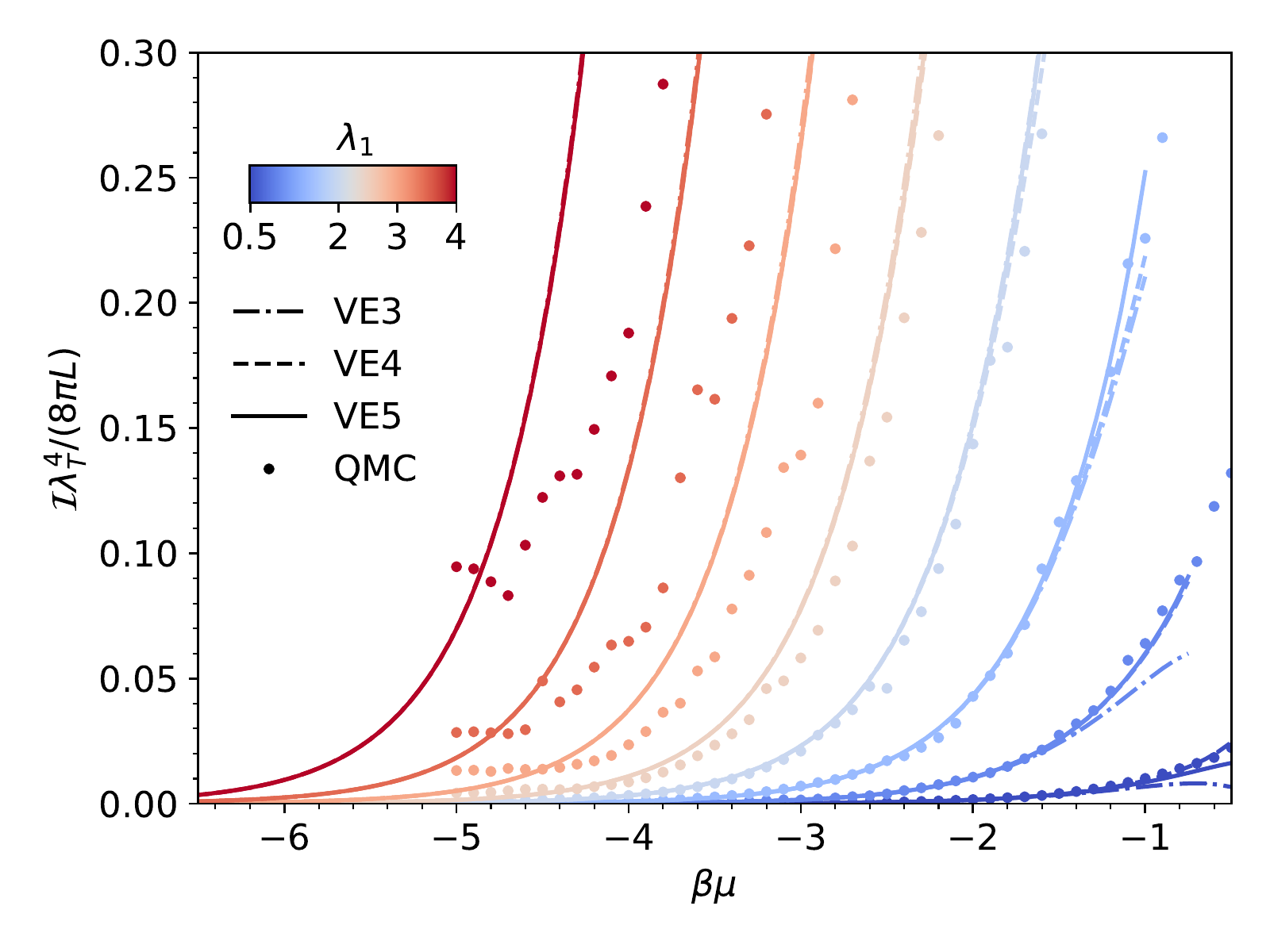}
  \end{center}
  \caption{Tan contact \( \mathcal{I} \), in the dimensionless form \( \mathcal{I} \pi \beta^2 / (2 L \lambda_1^2) \), as a function of \( \beta \mu \), for the attractive 1D Fermi gas.
    The color encodes the coupling strength of from \( \lambda_1 = 0.5 \) (dark blue) to \( \lambda_1 = 4.0 \) (dark red) in steps of \( 0.5 \).
    The dots show the QMC results of Ref.\cite{EoS1D} and the curves show the virial expansion at various orders,
    following the same line style as in Fig.~\ref{Fig:density-polarized-1d}.}
  \label{Fig:contact-1d}
\end{figure}

As elucidated by Tan and others~\cite{ShinaTan1,ShinaTan2,ShinaTan3}, in systems with short-range interactions the short-distance and high-frequency behavior of
correlation functions is captured by a single quantity called the contact, often denoted by $\mathcal I$. The contact also enters celebrated 
pressure-energy and other exact relations as well as sum rules~\cite{ZhangLeggett,Werner,BraatenPlatter1,BraatenPlatter2,BraatenPlatter3,SonThompson,TaylorRanderia,BraatenReview,Valiente1,Valiente2,WernerCastin, McKenneyDrut, ValientePastukhov} and is therefore a crucial piece of the thermodynamic puzzle that complements 
conventional quantities. Below, we compare our virial-expansion results for the contact in 1D with QMC data.

In 1D, the contact is given by
\beq
\mathcal I = \frac{2}{\beta} \frac{\partial (\beta \Omega) }{\partial a_0},
\eeq
such that its virial expansion becomes
\beq
\mathcal I = \frac{4\pi}{\lambda^3_T} Q_1 \sum_{m=2}^{\infty} c_m z^m,
\eeq
where $\lambda_T = \sqrt{2\pi \beta}$ and
\beq
c_m = \sqrt{\frac{\pi}{2}} \lambda_1^2\frac{\partial \Delta b_m}{\partial \lambda_1}.
\eeq
The Beth-Uhlenbeck formula yields 
%
\beq
c_2 = \frac{\lambda_1^2}{4} + \frac{\sqrt{\pi}}{8} e^{\lambda_1^2/4} \lambda_1^3 [1 + \text{erf}(\lambda_1/2)],
\eeq 
as first shown in Ref.~\cite{EoS1D}. Using the $\lambda_1$ dependence of our results for $\Delta b_n$, we 
obtained the virial expansion of $\mathcal I$ up to fifth order.
In Fig.~\ref{Fig:contact-1d} we show the dimensionless, intensive form of the contact, $\mathcal{I} \pi \beta^2 / (2 L \lambda_1^2)$ as a function of temperature in the virial expansion.
At weak couplings (\( \lambda_1 \leq 2.0 \)), the virial expansion shows good agreement with the QMC results of Ref.\cite{EoS1D}.
However, as the coupling strength increases, the disparity becomes significant, which is not completely unexpected as both methods face challenges 
in the strong coupling regime: the radius of convergence of the virial expansion can be very small, such that the expansion ceases to be useful even at 
very negative \( \beta \mu \); on the other hand, the lattice QMC method may suffer from increasing lattice-spacing effects at strong coupling, due to the
reduced size of the two-body bound state.

\subsection{\label{Sec:ResultsE}Virial coefficients in 2D}
\begin{figure}[t]
  \begin{center}
  \includegraphics[scale=0.57]{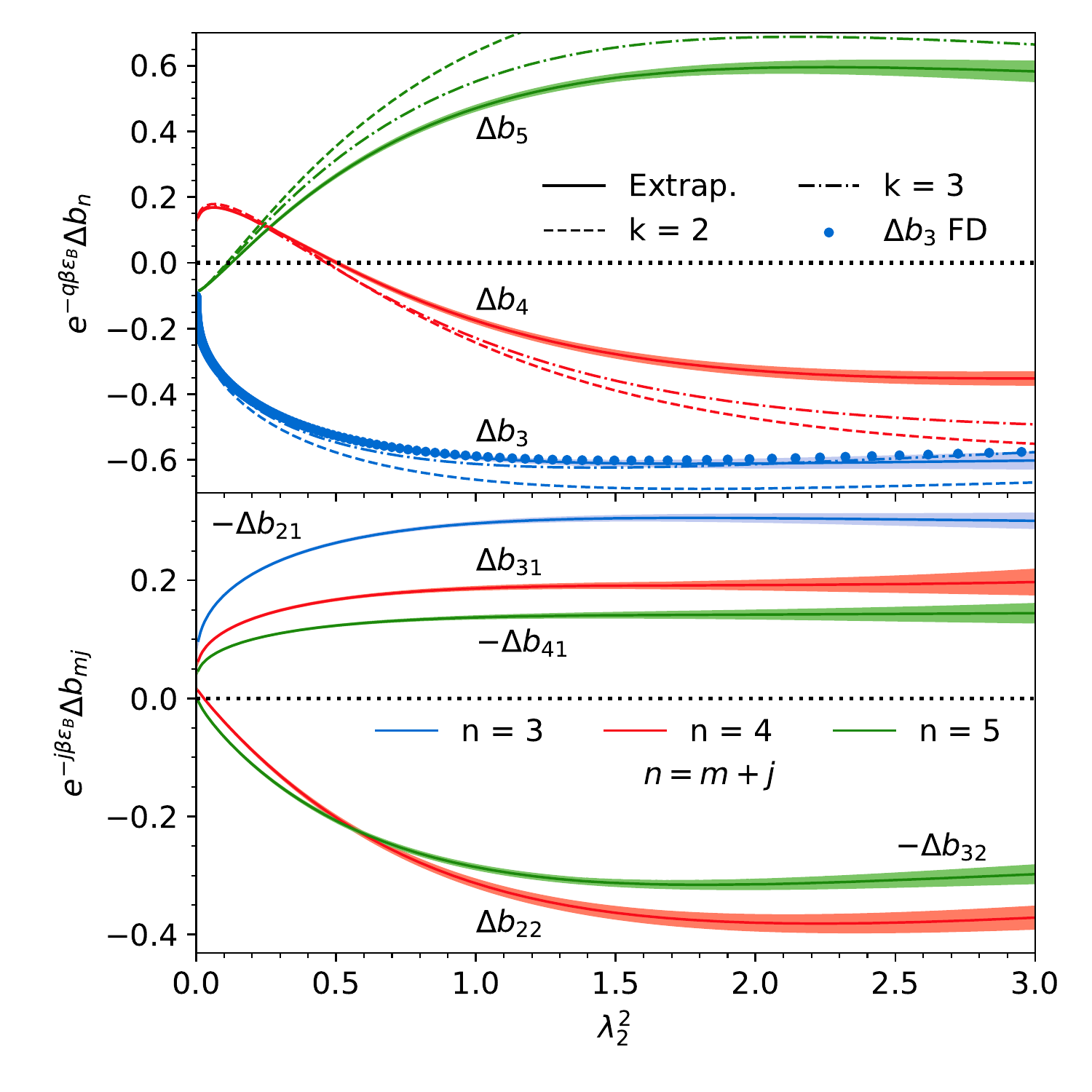}
  \end{center}
  \caption{{\bf Top}: Virial coefficients $\Delta b_n$ for $n=2-5$ for the 2D attractive Fermi gas, as a function of the square of the dimensionless coupling $\lambda_2$.
    To fit the scale, \( \Delta b_n \)s are scaled by \( \exp(-q \beta \epsilon_b)  \), where \( q \) is the maximum number of spin-$\uparrow\downarrow$ pairs,
    and \( \Delta b_{mj} \)s are scaled by \( \exp(-j \beta \epsilon_b) \).
    The $k=2$ results are shown as dashed lines, $k=3$ as dashed-dotted lines and the $k\to\infty$ extrapolation with solid lines.
    Results for $\Delta b_3$ appear in blue, for $\Delta b_4$ in red, and for $\Delta b_5$ in green.
   The diagrammatic result for $\Delta b_3$ from Ref.~\cite{virial2D2} appear as blue dots.
   {\bf Bottom}: Subspace contributions \( \Delta b_{mj} \) as functions of the coupling strength.
 Our results are shown as error bands, color-coded as in the top plot by $n = m+j$: 
 blue for $-\Delta b_{21}$, red for $\Delta b_{31}$ and $\Delta b_{22}$, and green for 
 $-\Delta b_{41}$ and $-\Delta b_{32}$.
   }
  \label{Fig:2Dbn}
\end{figure}
In Fig.~\ref{Fig:2Dbn} we show our results for $\Delta b_n$ for the 2D Fermi gas with attractive 
interactions~\cite{Theory2D}. As in 1D, we scaled the coefficients by $\exp(-q \beta \epsilon_B)$, where $q$
is the maximum number of available spin-$\uparrow\downarrow$ pairs. 
To renormalize, we again rely on the exact Beth-Uhlenbeck result~\cite{BU, DrummondVirial2D, virial2D, virial2D2, PhysRevA.89.013614, Ordo, Daza2D}
\beq
\Delta b_2^\text{2D,exact} =  
 e^{\lambda_2^2} - \int_0^{\infty} \frac{dy}{y} \frac{2 e^{-\lambda_2^2 y^2}}{\pi^2 + 4 \ln^2 y},
\eeq
to define $\tilde C$ as a function of the physical coupling $\lambda^2_2 = \beta \epsilon_B$,
where $\epsilon_B$ is the binding energy of the two-body system.

At all orders the similarity with 1D is clear:
$\Delta b_3$ remains negative for all the couplings studied, whereas
$\Delta b_4$ and $\Delta b_5$ change sign at strong enough coupling as a result of a competition
between subspaces $\Delta b_{m1}$ and $\Delta b_{n2}$ governed by the number of spin-$\uparrow\downarrow$ pairs available in each subspace.
While $\Delta b_4$ and $\Delta b_5$ are calculated here for the first time (thus furnishing a prediction), $\Delta b_3$ was 
calculated in Ref.~\cite{virial2D2}. The results of the latter are shown in Fig.~\ref{Fig:2Dbn} with blue dots, which agree remarkably
well with our answers.

The subspace contributions $\Delta b_{mj}$ are shown in the bottom panel of Fig.~\ref{Fig:2Dbn} and 
we again note clear parallels with the 1D case. Specifically, the subspace terms contributing to $\Delta b_4$ and $\Delta b_5$ 
enter with similar magnitudes but opposing signs, indicating that the final results for $\Delta b_4$ and $\Delta b_5$ result from
subtle coupling-dependent cancellations. Furthermore, $|\Delta b_{m1}|$ and $|\Delta b_{m2}|$ follow consistent trends as a 
function of the coupling across all $m$ (with the expected suppression as $m$ is increased). In fact, once the $\exp(-q \beta \epsilon_B)$ factor is included, $|\Delta b_{m1}|$ and $|\Delta b_{m2}|$ are approximately constant as the coupling is increased.

\begin{figure}[t]
  \begin{center}
  \includegraphics[scale=0.57]{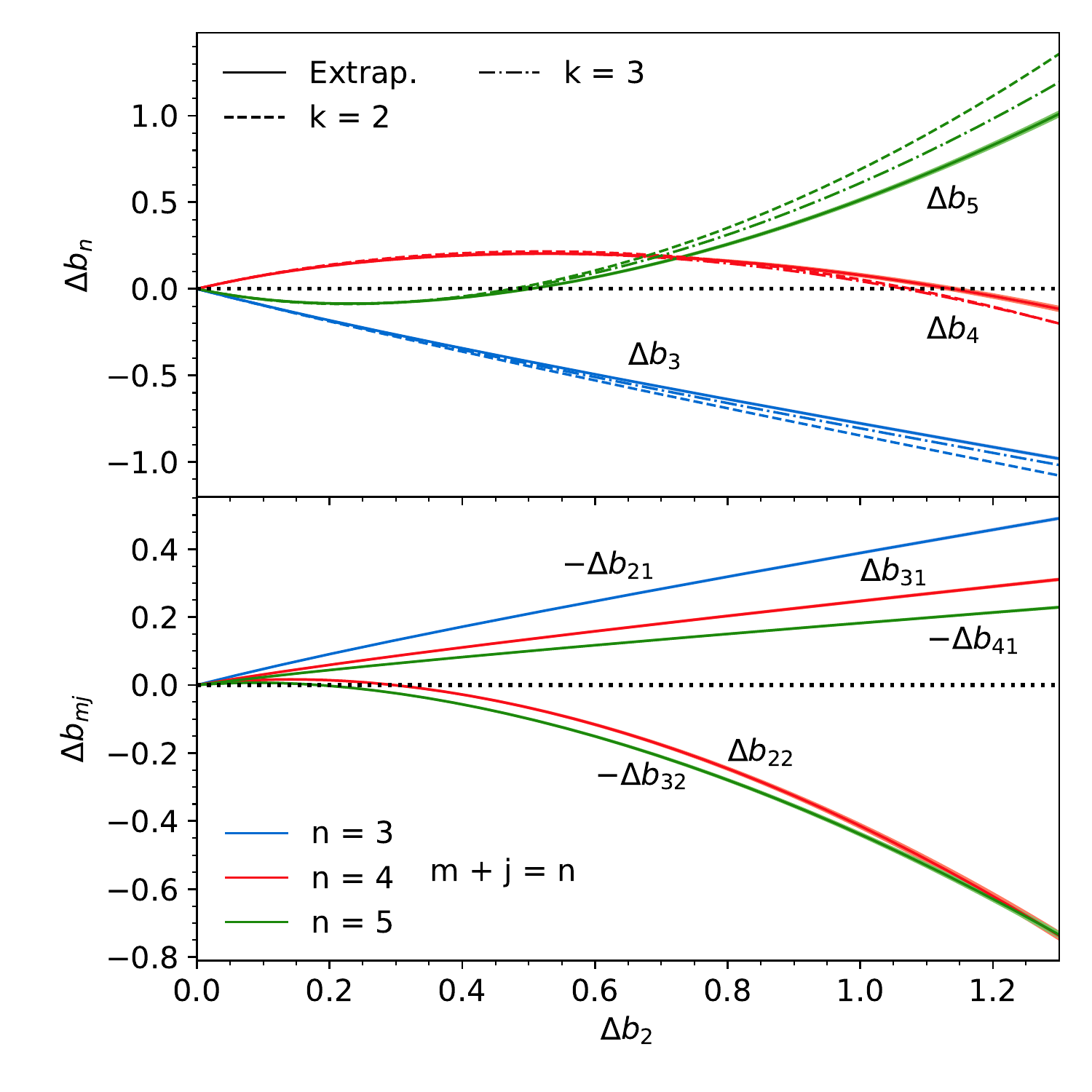}
  \end{center}
  \caption{Weak-coupling zoom-in on Fig.~\ref{Fig:2Dbn}, not including the $\exp(-q\beta \epsilon_B)$ factor and 
  plotted as a function of $\Delta b_2$. The right edge of the horizontal axis corresponds to $\lambda_2 \simeq 0.588$.
   }
  \label{Fig:2DbnZoom}
\end{figure}

At weak coupling, on the other hand, the details are elusive in the scale of Fig.~\ref{Fig:2Dbn}. There,
non-perturbative effects are only visible if shown differently, as we do in Fig.~\ref{Fig:2DbnZoom},
where show that, as functions of $\Delta b_2$ all the coefficients tend smoothly to zero as the coupling
is weakened. The non-perturbative behavior at weak coupling is completely captured
by $\Delta b_2$.
\begin{figure}[t]
  \begin{center}
   \includegraphics[scale=0.54]{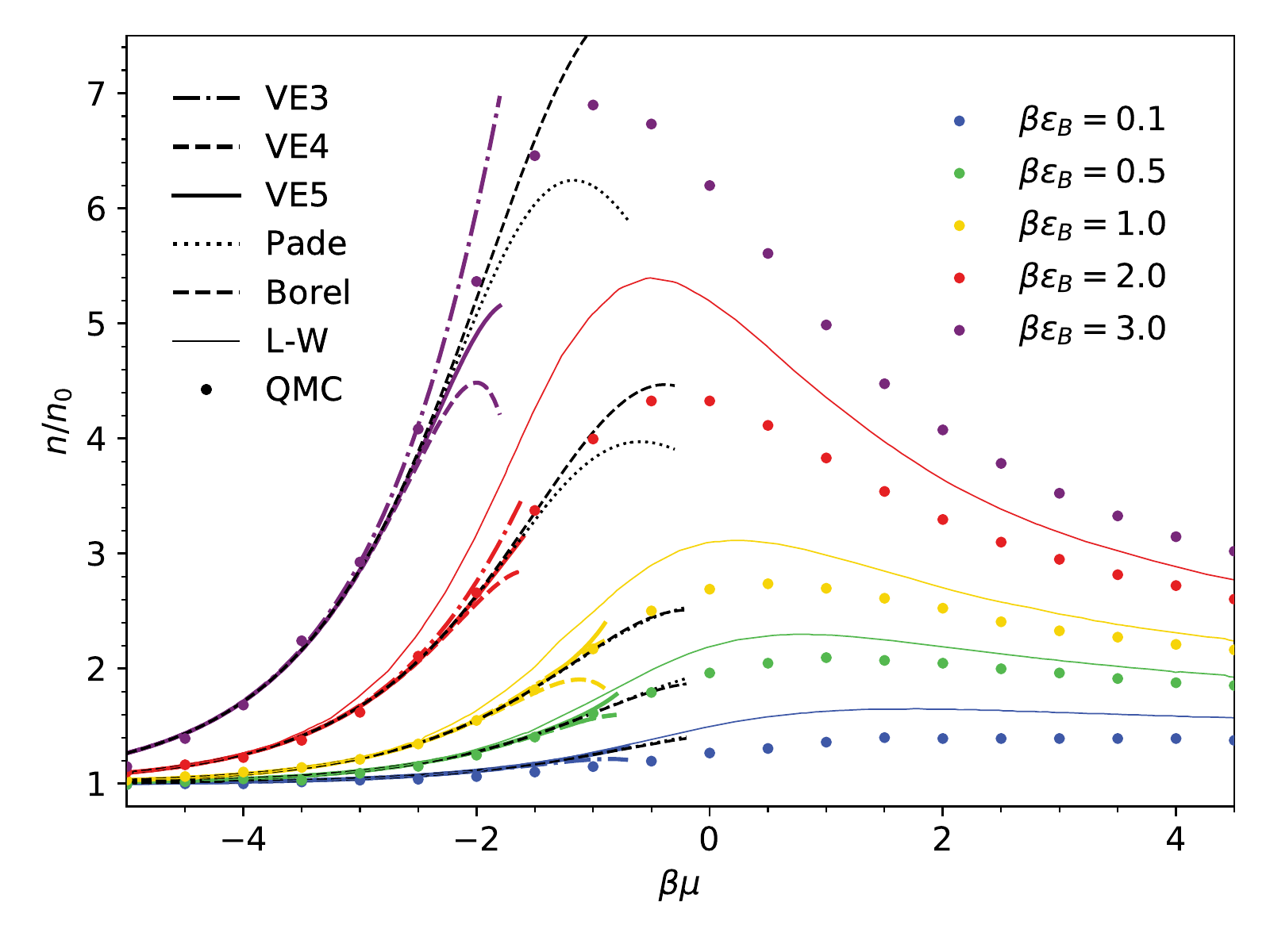}
  \end{center}
  \caption{Density equation of state \( n \), shown in units of the noninteracting counterpart \( n_0 \) at different coupling strength \( \lambda_2^2 = \beta \epsilon_B \) in 2D.
    The colored dots are the QMC results from Ref.~\cite{AndersonDrut}, the colored thin line shows the Luttinger-Ward result of Ref.~\cite{LWDensity2D},
    and the colored thick lines are virial expansions at different orders (same line style as in previous Fig.~\ref{Fig:density-polarized-1d}).
    The black dotted lines and dashed lines are the results, respectively, of Pad\'e and Borel resummation of order $[3/2]$ (see Sec.~\ref{Sec:ResumTech}) using the virial coefficients 
    up to the fifth order.
    }
  \label{Fig:density-2d}
\end{figure}
\begin{figure*}[t]
  \begin{center}
   \includegraphics[scale=0.39]{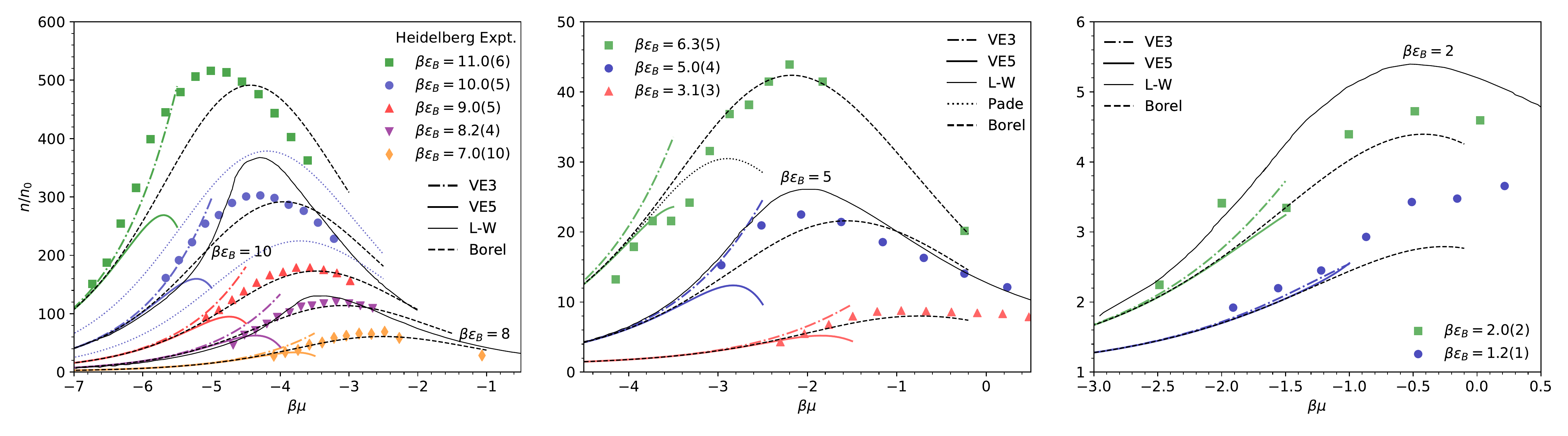}
  \end{center}
  \caption{
    Density equation of state \( n \) of the 2D attractive Fermi gas, shown in units of the noninteracting counterpart \( n_0 \) at different coupling strength 
    \( \lambda_2^2 = \beta \epsilon_B \) in 2D. The colored symbols are the results of experimental analyses from Ref.~\cite{Exp4} 
    and the colored lines show the virial expansion at different orders (following the same line style as in previous figures).
     [Note: the fourth-order case is omitted from this figure only for the clarity.]
    The black dashed lines are the results of Borel resummation of order $[3/2]$ (see Sec.~\ref{Sec:ResumTech}) using coefficients up to the fifth order.
    The thin black solid lines are the Luttinger-Ward results of Ref.~\cite{LWDensity2D}, around which the nearby labels indicate the corresponding \( \beta \epsilon_{B} \).
     Only the central value of our estimates is used in these plots; the relative-error on the virial coefficients from continuous-time 
     extrapolation can reach 15\% at the strongest couplings shown in the left panel, although that does not translate into a significant change
     in the scale shown. Also in the left panel, the blue dotted lines show our Borel-resummed results at \( \beta \epsilon_B = 9.5 \) (lower) and 10.5 (upper) respectively, which enclose the experimental points completely (in agreement with the experimental uncertainty on $\beta \epsilon_B$).
     In the middle panel, we also present the [3/2] Pad\'e result as a black dotted line for \( \beta \epsilon_B = 6.3 \) to demonstrate the performance difference between the two resummation techniques at strong coupling. Reference~\cite{Exp4} also provides data at \( \beta \epsilon_B = 0.45(5) \), but mostly at 
     large $\beta \mu > 0$ and therefore not shown here.
      }
  \label{Fig:density-2d-jochim}
\end{figure*}
\begin{figure}[t]
  \begin{center}
   \includegraphics[scale=0.54]{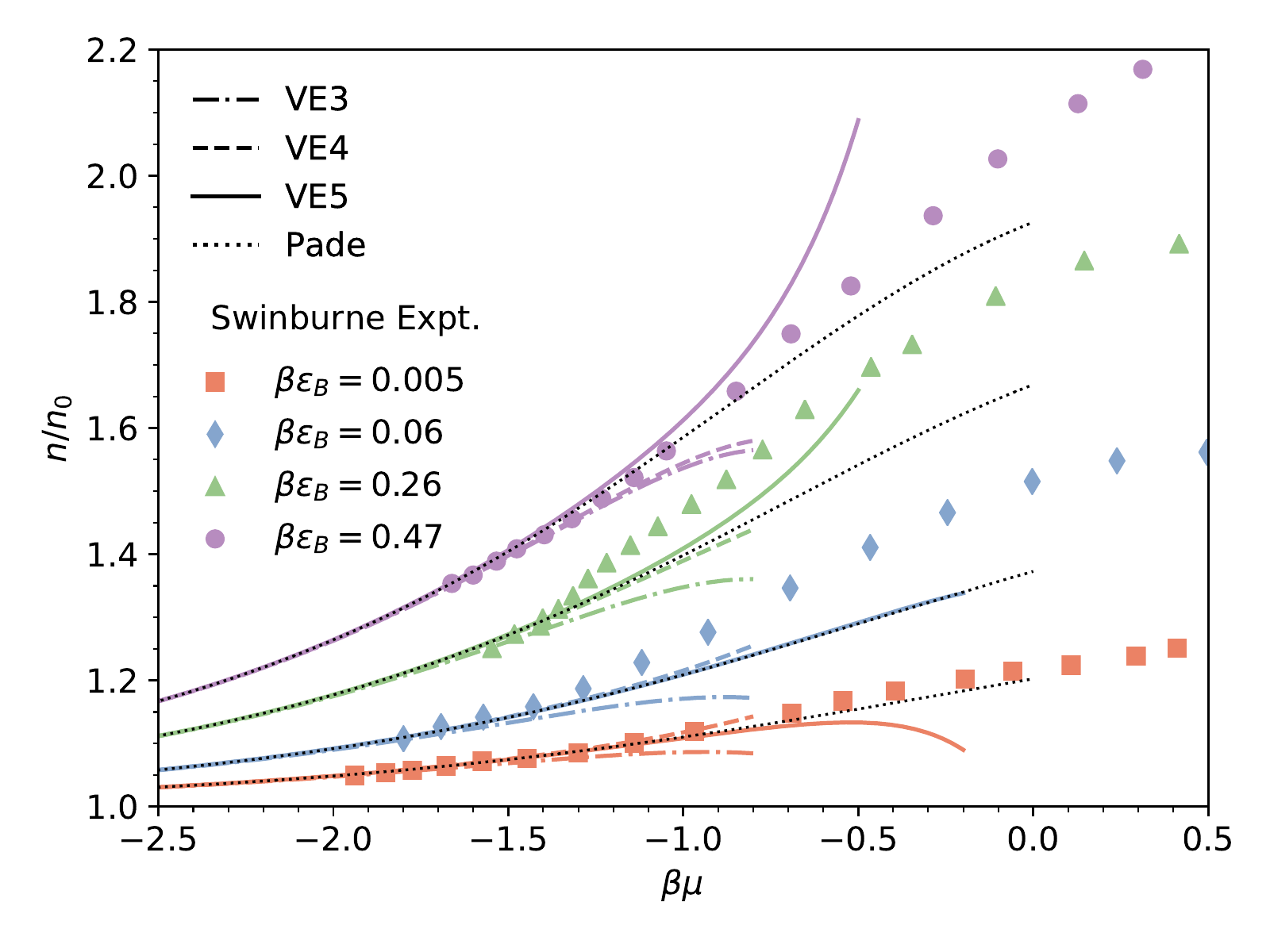}
  \end{center}
  \caption{Density equation of state \( n \), shown in units of the noninteracting counterpart \( n_0 \) at different coupling strength \( \lambda_2^2 = \beta \epsilon_B \) in 2D.
    The colored symbols are the results of experimental analyses from Ref.~\cite{Exp3} and the colored lines are virial expansions at different orders (the same line style applied).
    The black dotted lines and dashed lines are the results, respectively, of Pad\'e and Borel resummation (see Sec.~\ref{Sec:ResumTech}) of order $[3/2]$ using coefficients up to the fifth order.
    }
  \label{Fig:density-2d-vale}
\end{figure}
\begin{figure}[t]
  \begin{center}
   \includegraphics[scale=0.54]{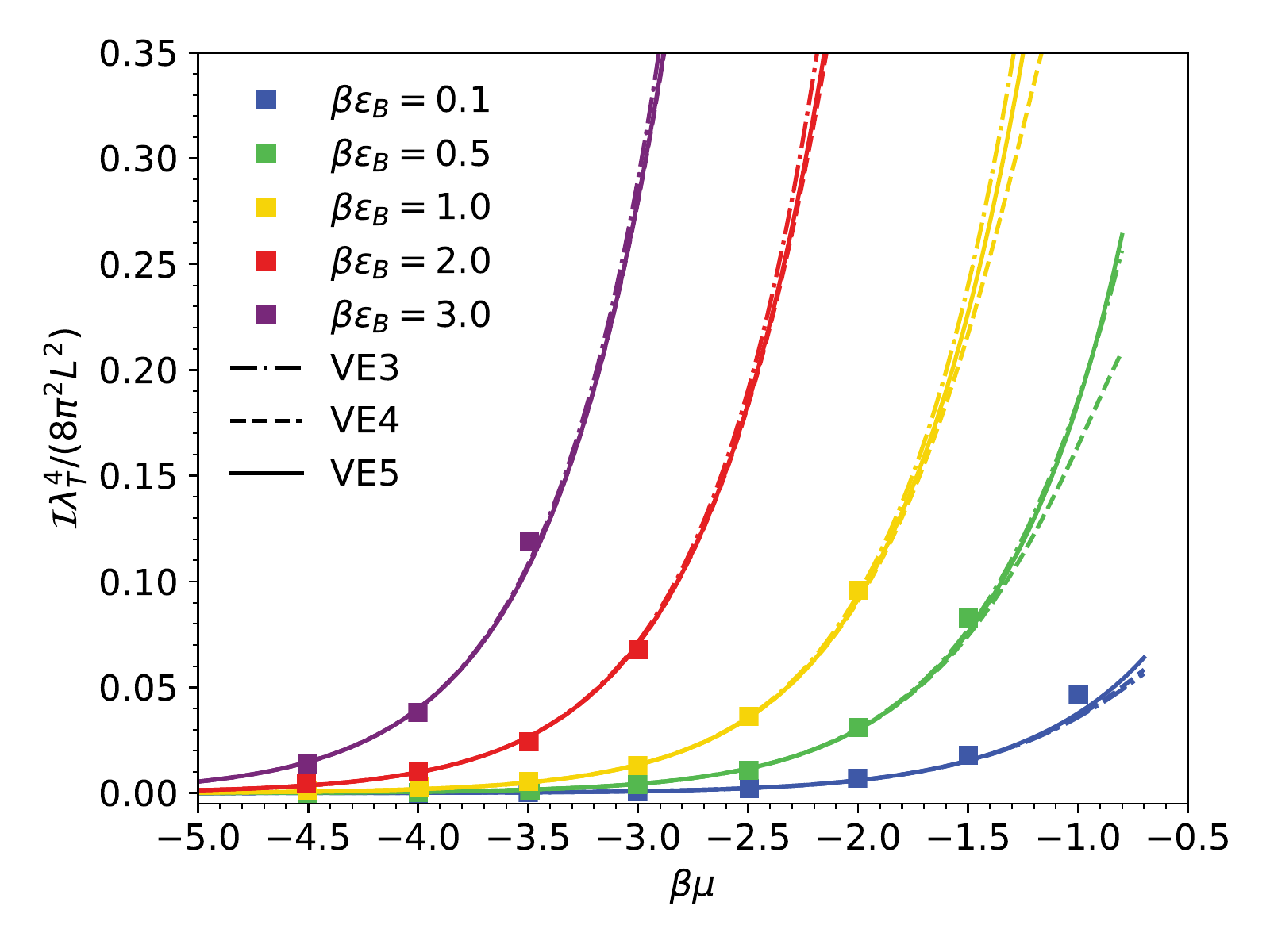}
  \end{center}
  \caption{Contact \( \mathcal{I} \), represented in the dimensionless form \( \mathcal{I} \lambda_T^4 / (8 \pi^2 L^2) \), as a function of \( \beta \mu \) in 2D.
    The colored squares are the QMC results of Ref.~\cite{AndersonDrut} and the colored lines are the results of virial expansion at various orders.
    }
  \label{Fig:contact-2d}
\end{figure}
%

\subsection{\label{Sec:ResultsF}Applications in 2D}

\subsubsection{Density equation of state at zero polarization}

Equation~(\ref{Eq:DensityUp}) can be easily applied in the unpolarized case 
by setting \( z_{\uparrow} = z_{\downarrow} \), such that the total density is given by
\begin{equation}
\lambda_T^d n  = \lambda_T^{d} n_0  + 2\sum_{n=2}^{\infty} n \Delta b_n z^{n},
\end{equation}
where the factor of $2$ accounts for the number of species.
Our results for $\Delta b_{n}$ yield the curves shown in Fig.~\ref{Fig:density-2d}, where they are compared with the QMC results of 
Ref.~\cite{AndersonDrut} and the Luttinger-Ward results of Ref.~\cite{LWDensity2D}. The agreement with the QMC data is
outstanding. Moreover, for all the couplings in the figure the Pad\'e and Borel resummations substantially extend the region of agreement.
With the exception of the weakest coupling in the figure (where the QMC results likely incur volume effects due to the large size of the two-body 
bound state), the virial expansion is systematically closer to the QMC data than to the Luttinger-Ward results. Although small, this discrepancy 
remains unresolved.

In Figs.~\ref{Fig:density-2d-jochim} and~\ref{Fig:density-2d-vale}, we compare our results
with the experimental data of Refs.~\cite{Exp4} and~\cite{Exp3}, respectively. In all cases, the agreement is remarkable in the regions
where the virial expansion is expected to work. Naturally, that region is pushed to progressively more negative $\beta \mu$ as the coupling
is increased; in other words, the radius of convergence of the virial expansion decreases as the coupling increases. 
However, it is also clear that, beyond weak and intermediate couplings (roughly up to $\beta \epsilon_B = 3.0$), the
benefits of pushing the virial expansion up to fifth order start to diminish, if the virial expansion is taken at face value. We find,
on the other hand, that Pad\'e and Borel resummations dramatically enhance the usefulness of the virial coefficients. As shown in 
Fig.~\ref{Fig:density-2d-jochim} (left and center panels in particular), our Borel resummations of the virial expansion 
agree not only qualitatively but in several cases also quantitatively with the experimental data.

\subsubsection{Contact}

In 2D, the contact is defined as
\beq
\mathcal I = \frac{2\pi}{\beta}\frac{\partial (\beta \Omega) }{\partial \ln(a_0/\lambda_T)},
\eeq
such that its virial expansion becomes
\beq
\mathcal I = \frac{(2\pi)^2}{\lambda_T^2} Q_1 \sum_{m=2}^{\infty} c_m z^m,
\eeq
where
\beq
c_m =\lambda_2 \frac{\partial \Delta b_m}{\partial \lambda_2}.
\eeq
The Beth-Uhlenbeck formula yields 
\beq
c_2 = 2 \lambda_2^2 e^{\lambda_2^2} 
\left[ 1 + 2 \int_0^{\infty} \dd y \frac{y e^{-\lambda_2^2(y^2 + 1)}}{\pi^2 + 4 \ln^2 y}
\right],
\eeq 
as shown in Refs.~\cite{virial2D2, AndersonDrut}. Using the $\lambda_2$ dependence of our results for $\Delta b_n$, 
we obtained the virial expansion of $\mathcal I$ up to fifth order. In Fig.~\ref{Fig:contact-2d} we show the dimensionless, 
intensive form of the contact, \( \mathcal{I} \lambda_T^4 / (8 \pi^2 L^2) \), as a function of \( \beta \mu \) in the virial expansion
and compared with the QMC results from Ref.~\cite{AndersonDrut}. As with the equation of state shown above, the 
agreement with the QMC data is remarkable in the region where the virial expansion is expected to work well.

\subsection{\label{Sec:ResultsG}Virial coefficients in 3D}
Finally, in 3D we have~\cite{BU, LeeSchaeferPRC1}
\bea
\Delta b_2^\text{3D,exact} = \frac{e^{\lambda_3^2}}{\sqrt{2}}[1 + \mbox{erf}(\lambda_3)],
\eea
where $\lambda_3 = \sqrt{\beta}/a_0$, and $a_0$ is the s-wave scattering length.
Note the unitary limit corresponds to $\lambda_3 = 0$
and we only explore \( \lambda_3 < 0 \) regime in this work.

\begin{figure}[t]
  \begin{center}
  \includegraphics[scale=0.56]{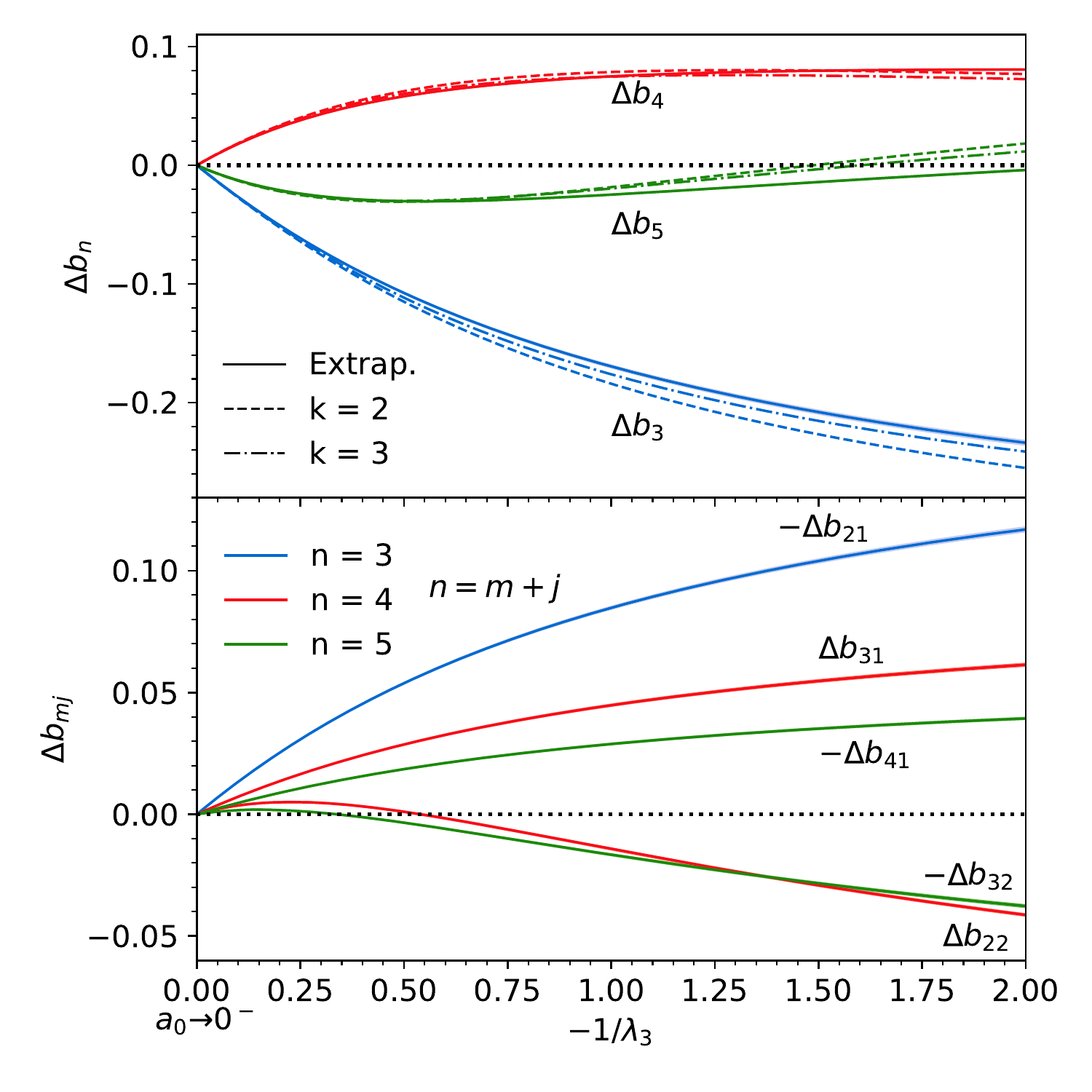}
  \end{center}
  \caption{$\Delta b_n$ for $n=2-5$ for the 3D Fermi gas in the weak-coupling regime, as a function of the dimensionless coupling $\lambda_3$.
  The $k=2$ results are shown with dashed lines,  
   $k=3$ with dashed-dotted lines, and the $k\to\infty$ extrapolation with solid lines. Results for $\Delta b_3$ appear in blue, for $\Delta b_4$ in red, and for $\Delta b_5$ in green.
   }
  \label{Fig:3DbnWeak}
\end{figure}

\begin{figure}[t]
  \begin{center}
  \includegraphics[scale=0.56]{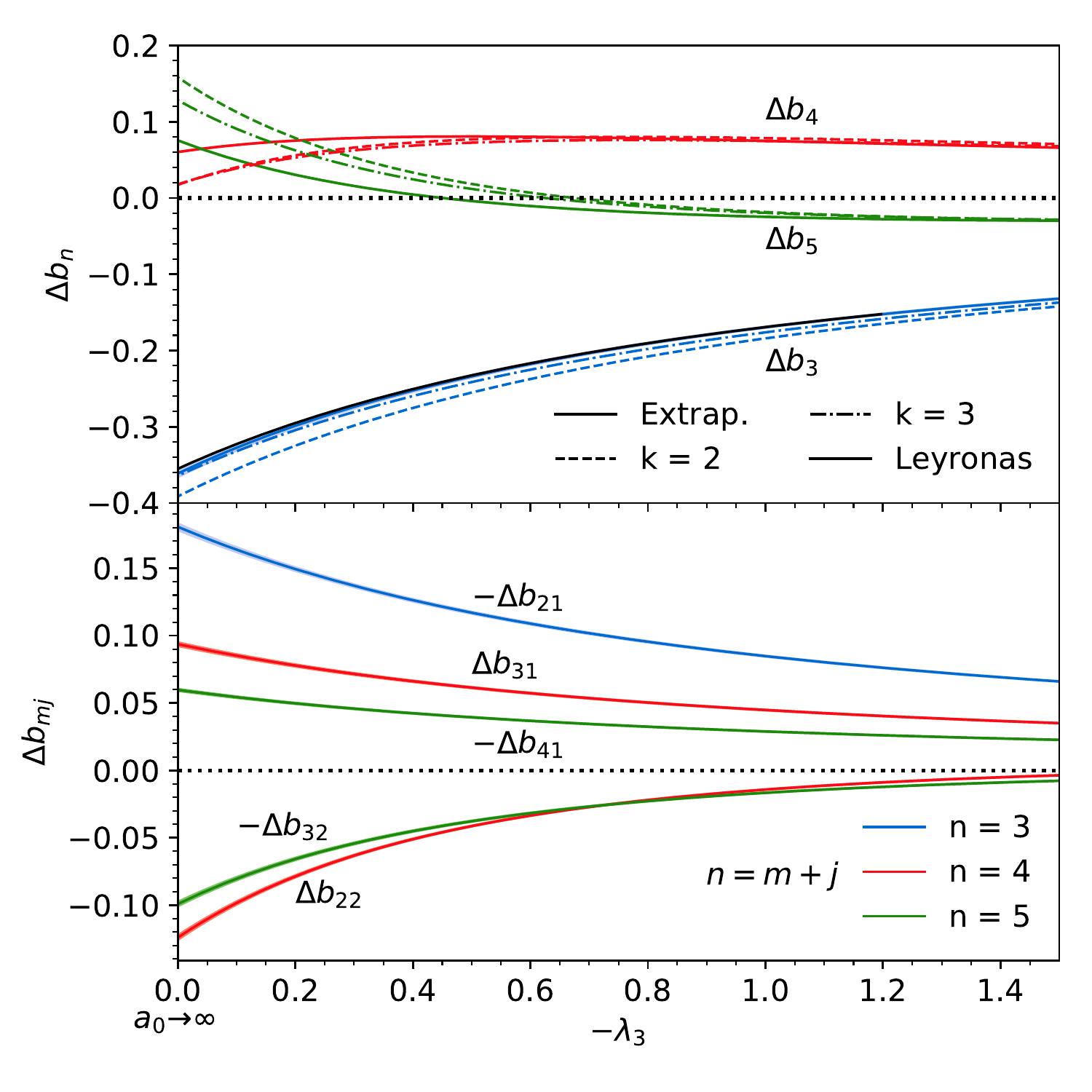}
  \end{center}
  \caption{$\Delta b_n$ for $n=2-5$ for the 3D Fermi gas in the strong-coupling regime, as a function of the dimensionless coupling $\lambda_3$.
   [Note that the unitary limit corresponds to $\lambda_3 = 0$.]
   The exact result for $\Delta b_3$ from Ref.~\cite{Leyronas} appears as a thin solid black line.
  The $k=2$ results are shown with dashed lines,  
   $k=3$ with dashed-dotted lines, and the $k\to\infty$ extrapolation with solid lines. 
   Results for $\Delta b_3$ appear in blue, for $\Delta b_4$ in red, and for $\Delta b_5$ in green.
   }
  \label{Fig:3DbnStrong}
\end{figure}

In Fig.~\ref{Fig:3DbnWeak} and Fig.~\ref{Fig:3DbnStrong}, we present our results for $\Delta b_n$ for the 3D Fermi gas with attractive interactions.
These plots parallel the main figure of our previous work of Ref.~\cite{HouDrut}, where we show the same results as a function
of $\Delta b_2$. For completeness, future reference, and to parallel our discussion in lower dimensions, 
Fig.~\ref{Fig:3DbnWeak} shows our results as a function of $-1/\lambda_3$ to display the weak coupling regime, while
in Fig.~\ref{Fig:3DbnStrong} we plot the approach to the unitary limit ($\lambda_3 = 0$) as a function of $\lambda_3$.
In the latter, we find excellent agreement with the exact $\Delta b_3$ of Ref.~\cite{Leyronas} (see also 
Refs.~\cite{LiuHuDrummond, DBK, Rakshit, Ngampruetikorn}).

The bottom panels of Fig.~\ref{Fig:3DbnWeak} and~\ref{Fig:3DbnStrong} show the subspace decomposition of $\Delta b_n$
into $\Delta b_{mj}$. The qualitative similarities with lower dimensions are clear. Here, however, we have explored a coupling regime in
which no two-body bound states are yet formed, such that the introduction of the $\exp(-q\beta\epsilon_B)$ factor is not needed.
As the unitary limit is approached, however (see left edge of Fig.~\ref{Fig:3DbnStrong}), all of the coefficients we computed start
to display increased curvature. Beyond that point, we expect an exponential increase characterized by $\exp(q\beta\epsilon_B)$ 
precisely as in lower dimensions. The rapid downturn of $\Delta b_3$ at strong coupling was, in fact, already noticed in Ref.~\cite{Leyronas}, 
while the same feature for $\Delta b_4$ appeared in Ref.~\cite{Ngampruetikorn}. Here, we see that that behavior is inevitable as it
is merely a consequence of the dominance of $\Delta b_{m2}$ (with $m = 2,3$) over $\Delta b_{m1}$ (with $m = 3,4$). However, 
the sign difference between these terms, as in lower dimensions, results in cancellations that make fully numerical determinations of
$\Delta b_4$ and $\Delta b_5$ a difficult task. Finally, we note that our analysis of $\Delta b_{mj}$, and their apparent systematic 
behavior, suggests that the conjecture of Ref.~\cite{Bhaduri1,Bhaduri2} on the high-order virial coefficients in the unitary limit may be refined 
by focusing on the subspaces rather than the full $\Delta b_{n}$.

%

\subsection{\label{Sec:ResultsH}Applications in 3D}

\begin{figure}[t]
  \begin{center}
   \includegraphics[scale=0.53]{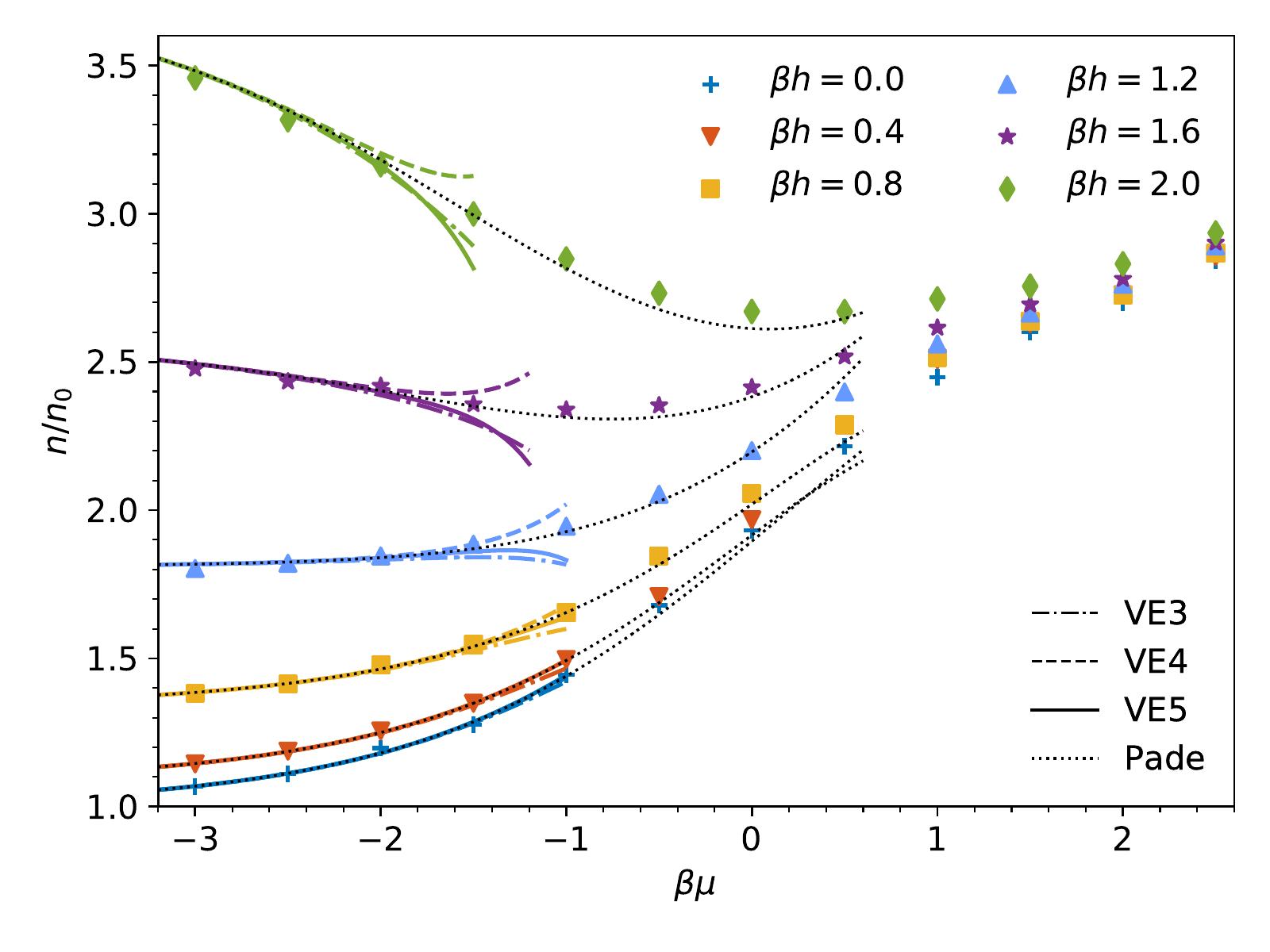}
  \end{center}
  \caption{Density equation of state \( n \) of the unitary Fermi gas, shown in units of the noninteracting, unpolarized counterpart \( n_0 \), for several values of the chemical potential difference $\beta h = \beta (\mu_\uparrow - \mu_\downarrow)/2$.
    The colored symbols are complex-Langevin results from Ref.~\cite{polarizedUFGCL} and the colored lines show the virial expansion at 
    various orders: dashed-dotted line at third order, dashed line at fourth order, and solid line at fifth order.
    The black dotted line is the result of the $[3/2]$ Pad\'e resummation (see Sec.~\ref{Sec:ResumTech}).
    The limiting value for $\beta \mu \to -\infty$ is known exactly and is given by $\cosh(\beta h)$.
  }
  \label{Fig:density-polarized-3d}
\end{figure}
%

\subsubsection{Density equation of state in the unitary limit}

In Fig.~\ref{Fig:density-polarized-3d}, we present our results for the density equation of state
in the unitary limit and compare with the complex-Langevin results of Ref.~\cite{polarizedUFGCL}.
The fifth-order expansion shows the best agreement compared to its lower-order counterparts, although
the improvement is mostly marginal. When applying the Pad\'e resummation technique, however,
the agreement with the data is extended even beyond $\beta \mu=0$. Notably, the change in curvature displayed
by the data is reproduced by the Pad\'e approximant. Beyond $\beta \mu = 0.6$, however, the Pad\'e approximant 
progressively departs from the data. The Borel-Pad\'e resummation shows performance very similar to that of pure 
Pad\'e resummation and is therefore omitted.
\begin{figure}[t]
\begin{center}
\includegraphics[scale=0.53]{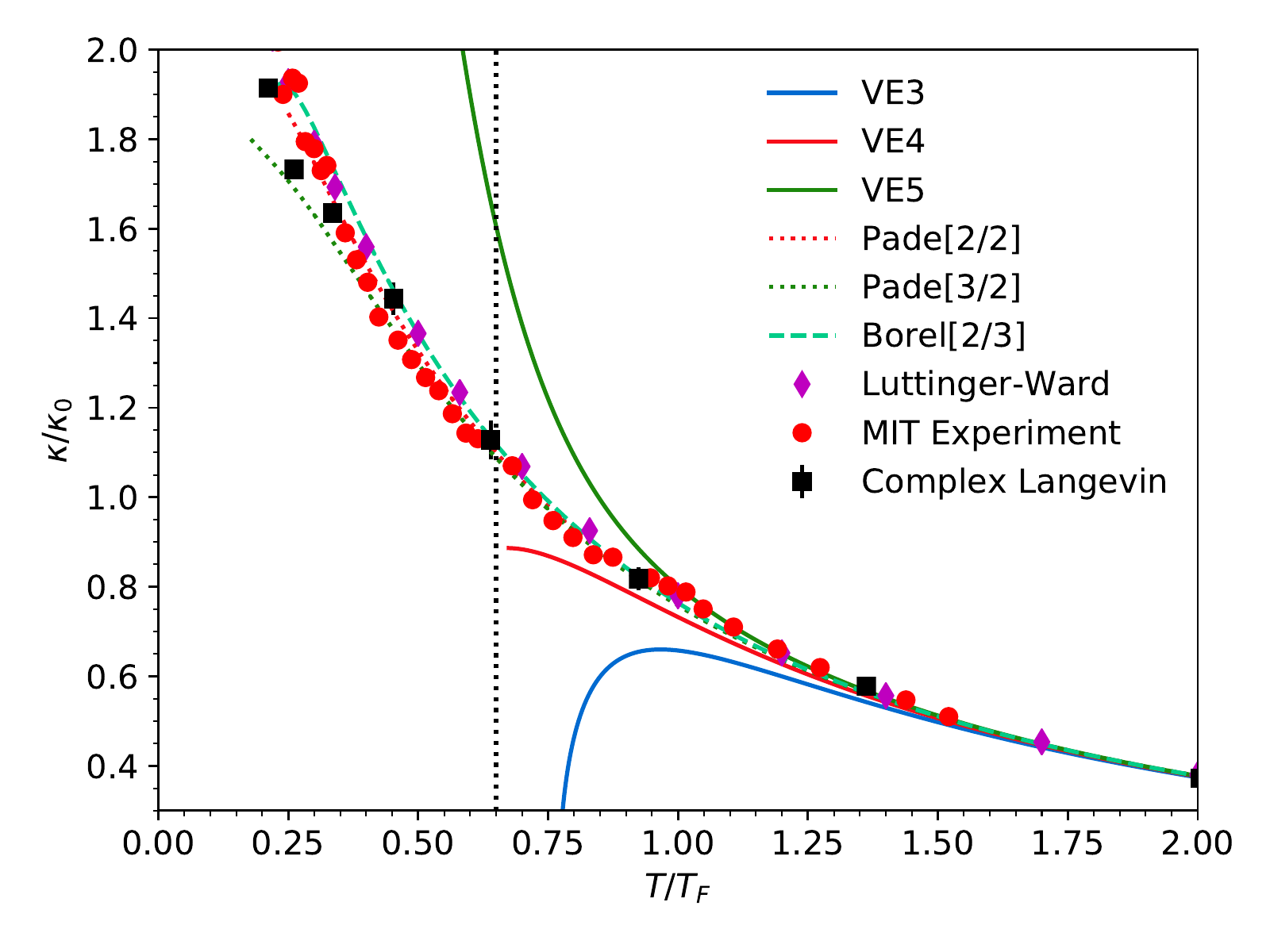}
\end{center}
\caption{Compressibility \( \kappa \) of the unitary Fermi gas in units of its noninteracting, ground-state counterpart \( \kappa_0 \),
as a function of the temperature $T$ in units of the Fermi temperature $T_F$.
Our results for the virial expansion for fugacity \( z \in [0, 1] \) are shown as solid lines:
blue for third order, red for fourth order, and green for fifth order.
The dotted lines are the result of Pad\'e resummation of order [2/2] (red) and [3/2] (green), respectively.
The cyan dashed line is the result of a [2/3] Borel-Pad\'e resummation (see Sec.~\ref{Sec:ResumTech}).
Other resummation orders are omitted due to the appearance of poles on the real axis in the region of interest.
The vertical dotted line corresponds to the approximate \( T/T_F \) where \( z = 1 \) for the three resummed results.
The red circles show the data from the MIT experiment of Ref.~\cite{Exp2}.
The purple diamonds are the Luttinger-Ward calculations of Ref.~\cite{EnssHaussmann}.
The black squares are complex-Langevin results from Ref.~\cite{polarizedUFGCL}.
\label{Fig:compressibility-3d}
}
\end{figure}
\subsubsection{Compressibility in the unitary limit}
The compressibility \( \kappa \) is defined as
\beq
    \kappa = -\frac{1}{V} \left[\pdv{V}{P}\right]_{T} =  \frac{\beta}{n^2} \left[\pdv{n}{\beta \mu}\right]_{T},
\eeq
where, in terms of virial expansion,
\beq
     \pdv{n}{(\beta \mu)} = \pdv{n_0}{(\beta \mu)} + \sum_{m=2}^{\infty} m^2 \sum_{i + j = m} \Delta b_{ij} z_{\uparrow}^i z_{\downarrow}^j.
\eeq
In Fig.~\ref{Fig:compressibility-3d}, we present our estimates for the compressibility \( \kappa \), shown in units of its noninteracting counterpart \( \kappa_0 = 3 / (2 n \epsilon_F) \), in the unitary limit and compare them with the 
experimental measurements from Ref.~\cite{Exp2}, the Luttinger-Ward calculations of Ref.~\cite{EnssHaussmann},
and the complex-Langevin results from Ref.~\cite{polarizedUFGCL},

The results of Pad\'e and Borel-Pad\'e resummations show much better agreement with experimental data than the finite-order virial expansion, even well beyond the region of the virial expansion \( z \ll 1 \). 
Specifically, the resummations smoothly follow the trend of the experimental data up to fugacities as large as $z = 10$ (maximum fugacity 
shown in Fig.~\ref{Fig:compressibility-3d}), which is surprising considering that the superfluid transition occurs at $z \simeq 10$.

\subsubsection{Tan contact}
For completeness, we cite the equations for the contact and its virial expansion, as done above in 1D and 2D.
Our results for the contact at unitarity are presented elsewhere~\cite{HouDrut}.
In 3D, the contact is defined as
\beq
\mathcal I = -\frac{4\pi}{\beta} \frac{\partial (\beta \Omega) }{\partial a_0^{-1}},
\eeq
such that its virial expansion reads
\beq
\mathcal I =  \frac{8\pi^2}{\lambda_T}  Q_1 \sum_{m=2}^{\infty} c_m z^m,
\eeq
where
\beq
c_m = \frac{1}{\sqrt{2\pi}}\frac{\partial \Delta b_m}{\partial \lambda_3}.
\eeq
The Beth-Uhlenbeck formula yields 
\beq
c_2 = \frac{1}{\pi} + \frac{1}{\sqrt{\pi}} \lambda_3 \; e^{\lambda_3^2} (1 + \text{erf}(\lambda_3)).
\eeq 
\\

\subsection{\label{Sec:ResumTech}Resummation techniques}
  Before concluding, we present more details on the resummation techniques used in this work, namely the Pad\'e and 
  Borel-Pad\'e resummations, which we have found useful in extending the applicability of the virial expansion.
  Generally speaking, the Pad\'e resummation, based on fitting a Pad\'e approximant (see below), is useful when a series
  is finitely truncated, as in the case of virial expansion.
  In such cases, the Pad\'e approximant is often better behaved than
  the partial sums of the original series, and it may even work where the original series diverges, i.e. 
  beyond the radius of convergence of the series.
  In our case, the technique is simple to apply: the coefficients $\Delta b_n$ that we calculated determine the
  unknown coefficients of a Pad\'e approximant which is a rational function $P(z)/Q(z)$, where $P$ and $Q$ are
  polynomials of degrees $a$ and $b$ respectively. Such an approximant is denoted by $[a/b]$.
  Using $m$ available virial coefficients, it is possible to fully determine $m$ coefficients in $P$ and $Q$, which means that 
  $a + b = m$ (note that the independent term in $Q$ is set to 1 by convention). By definition, the Taylor expansion of 
  $P(z)/Q(z)$ at small $z$ reproduces the input virial coefficients.
  
  The Borel-Pad\'e resummation is based on applying a Borel summation~\cite{BorelSum1, BorelSum2}, followed by a Pad\'e fit and reverse (integral) Borel
  transform (which is a Laplace transform). In short, the Borel summation amounts to replacing each coefficient $\Delta b_n$ by $\Delta b_n/n!$; the resulting function is
  fit with a Pad\'e approximant, and that approximant is then numerically integrated to (effectively) undo the introduction of the $n!$ factor.
  Further details on this well-known technique can be found, for instance, in the Supplemental Materials of Ref.~\cite{HouDrut}.
  
    Such an approach has been applied in many other areas (famously in the calculation of critical exponents~\cite{GuidaJustin}) ranging from QCD~\cite{QCDBorelPade} to ultracold atomic physics~\cite{NishidaSon, ArnoldDrutSon}. The method is particularly useful for summing diverging asymptotic series, which is the reason we used it in 2D at strong coupling.
  In fact, our empirical observations are that the Borel-Pad\'e resummation yields better qualitative behavior and better overall agreement with 
  experiments at strong coupling than the pure Pad\'e fit mentioned above (see middle panel of Fig.~\ref{Fig:density-2d-jochim}).  At weak coupling, on the other hand, both techniques yield similar results 
  (see, e.g., Fig.~\ref{Fig:density-2d}).
    In 3D, the strongest {\it effective} coupling corresponds to the unitary limit;
    beyond that point the strong attraction leads to the formation of bosonic dimers whose interaction is, as a residual effect, weakly repulsive.
    However, the two resummation methods still yield similar results at unitarity as the magnitude of the virial coefficients is small.
    It is only for positive \( \lambda_3 \) (corresponding to deeply bound pairs) that
    \( \Delta b_n \) will start to grow exponentially, as we saw in 1D and 2D, and in that case the Borel-Pad\'e resummation is expected to offer better estimates.
  
\begin{figure}[t]
  \begin{center}
   \includegraphics[scale=0.54]{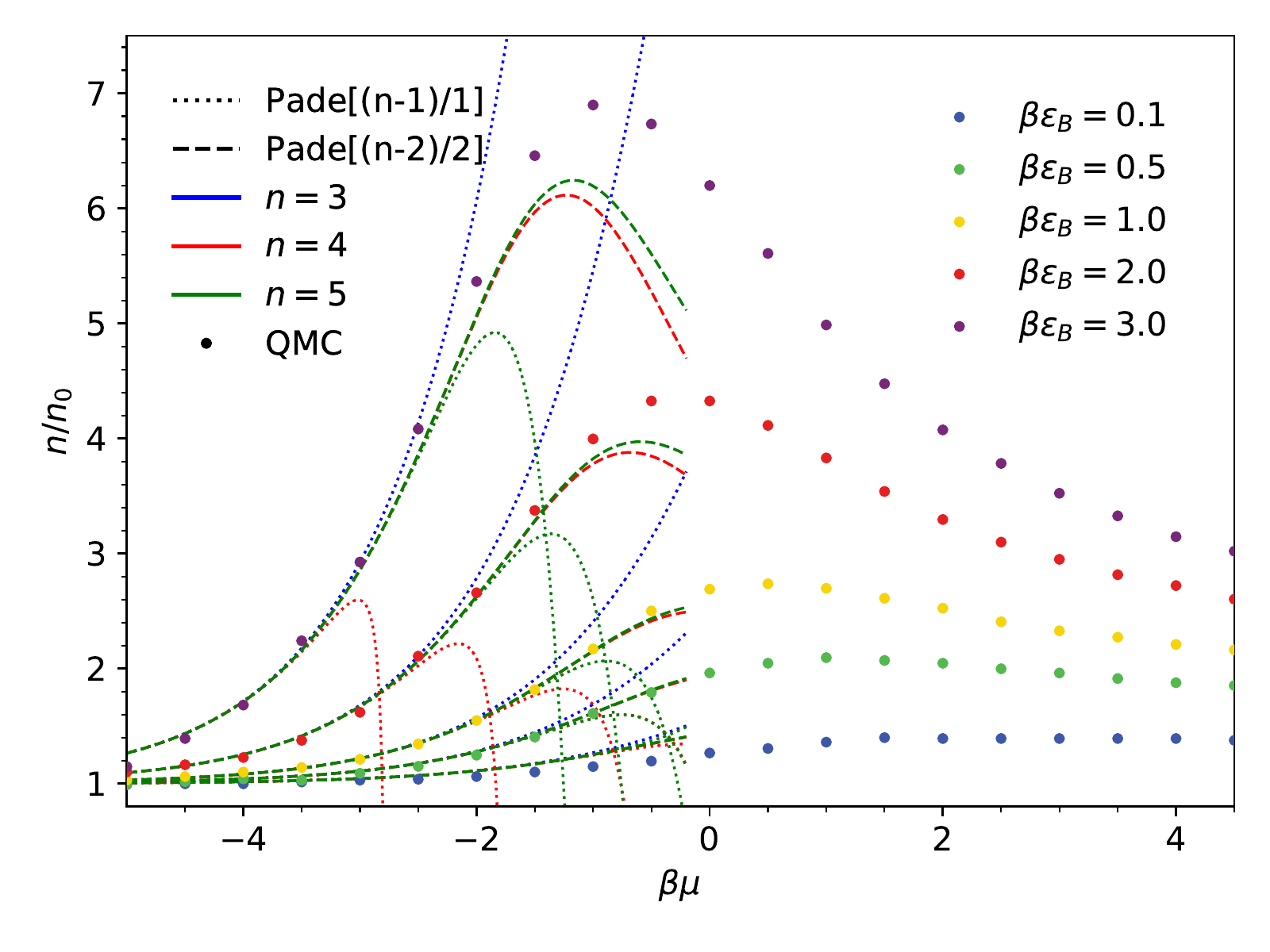}
  \end{center}
  \caption{Density equation of state \( n \), shown in units of the noninteracting counterpart \( n_0 \) at different coupling strength \( \lambda_2^2 = \beta \epsilon_B \) in 2D.
    The colored dots are the QMC results from Ref.~\cite{AndersonDrut}, the colored lines shows the results using Pad\'e approximation at different orders.
    Different colors indicate the higher order of coefficients used in calculating the Pad\'e approximant: blue for \( n = 3 \), red for \( n = 4 \) and green for \( n = 5 \);
    different linestyles represent the degree of denominator in the approximant: dotted line for \( [(n-1)/1] \) and dashed line for \( [(n-2)/2] \).
    }
  \label{Fig:density-2d-pade-comparison}
\end{figure}

  In Fig.~\ref{Fig:density-2d-pade-comparison}, we show in more detail a comparison of Pad\'e and Pad\'e-Borel resummations at different orders, as applied to the density equation of state of the 2D attractive Fermi gas.
  For all couplings, the pure Pad\'e approach of order $[3/2]$ shows the best agreement with the QMC calculations compared with every other 
  lower-order approximant.[For the $[1/4]$ approximant we have found no solution for the Pad\'e coefficients f
  or the given values of the virial coefficients.] 
  We find that, for \( \beta \epsilon_B = 2.0 \) and \( 3.0 \), the $[3/1]$ approximant encounters poles on the positive $z$ axis at 
  \( \beta \mu \approx -2.7 \) and \( -1.7 \) respectively, which underscores the importance of calculating high-order virial coefficients in
  order to access a wide range of possible Pad\'e approximants.
  
Although we found remarkable agreement between our resummations and other results in several examples, these techniques admittedly 
represent a break from the main {\it a priori} method we used to obtain the virial coefficients, as Pad\'e and Borel methods have not 
been rigorously justified here (to the best of our knowledge). Future studies may shed light on this matter.
%

\section{\label{Sec:Conclusion}Summary and Conclusions}

In this work we calculated the interaction-induced change in the third to fifth virial coefficients,
$\Delta b_3$ -- $\Delta b_5$, of spin-$1/2$ fermions with attractive interactions using a 
temporal lattice approximation. We provided a few analytic answers in coarse discretizations $k=1,2$
before pushing our results to high $k$ ($k=21$ for $\Delta b_3$; $k=12$ for $\Delta b_4$; and
$k=9$ for $\Delta b_5$) and extrapolating to the continuous-time limit.
Using a renormalization prescription based on matching $\Delta b_2$ to the known exact results,
we obtained $\Delta b_n$ for $n = 3,4,5$ for a range of attractive couplings in 1D, 2D, and 3D.

In 1D, our results for $\Delta b_n$ agree with previous QMC estimates (obtained at weak coupling) 
and substantially extend the coupling range. In addition, we obtained the subspace contributions which
as in higher dimensions, appear with similar magnitude but opposite sign, thus partially cancelling each
other out. We found that the presence of a two-body bound state strongly controls the magnitude
of the virial coefficients: multiplying by a factor $\exp(-q \beta \epsilon_B)$, where $q$
is the maximum number of $\uparrow,\downarrow$ pairs and $\epsilon_B$ their binding energy,
it is possible to show all the virial coefficients we calculated on the same vertical scale.

As an application in 1D, we compared our results (both partial sums as well as Pad\'e resummation) 
with complex Langevin results for the density equation of state at finite polarization, where we found
excellent agreement. Similarly, we compared with QMC data for the 
Tan contact in the unpolarized case, where we found very good agreement at weak coupling, deteriorating 
as the coupling is increased (likely due to the limited range of validity of the expansion, but also due
to the decreased quality of the data and possible lattice spacing effects at strong coupling).

In 2D, the exponential growth mentioned above for 1D is even more evident: all the coefficients we calculated 
become very approximately constant once the $\exp(-q \beta \epsilon_B)$ factor is included.
The well-known nonperturbative features of this system at weak coupling are smoothly captured
by plotting as a function of $\Delta b_2$ rather than $\lambda_2^2 = \beta \epsilon_B$.
Our results for $\Delta b_3$ match very closely those of Ref.~\cite{virial2D2}.

As an application in 2D, we compared our results with QMC and experimental data on the density equation of state
of the unpolarized system. Partial sums of the virial expansion and resummation results
are in excellent agreement with QMC data for a range of negative values of $\beta \mu$, as expected.
Furthermore, in all cases the fifth-order of the virial expansion yields an improvement over lower orders.
Beyond partial sums, the Pad\'e and Pad\'e-Borel resummation shows outstanding agreement in a range of $\beta \mu$ 
that is substantially larger than that of the partial sums. When comparing with experimental data, the picture
is similar: the partial sums give at least reasonable agreement where expected, but it is the Pad\'e-Borel resummation
that brings about the most remarkable overall agreement with the data. Finally, our comparison with the 
Tan contact obtained by QMC methods, where available, shows excellent agreement.

In 3D, we calculated $\Delta b_n$ for couplings up to the unitary limit. Our results for $\Delta b_3$ show remarkable 
agreement with the exact result of Ref.~\cite{Leyronas} at all couplings. For the unitary limit, our results
were discussed at length elsewhere~\cite{HouDrut}. As no bound states are formed for the couplings we explored
in 3D, there is no need to include an exponential factor as in the 1D and 2D cases; the growth of the virial 
coefficients is in fact relatively mild for all the couplings we explored in 3D.

As an application in 3D, we have calculated the density equation of state at finite chemical
potential asymmetry, as well as the compressibility of the unpolarized system. As in the 1D case, we find 
that the Pad\'e resummation substantially extends the usefulness of the virial expansion. In particular,
for the compressibility the agreement with experiments extends as far as $z=10$.

In 1D, 2D, and 3D we presented not only $\Delta b_n$ for $n=3,4,5$, but also the subspace contributions,
crucially $\Delta b_{m1}$ and $\Delta b_{m2}$. Discerning these is important because they determine
the thermodynamics of the polarized version of the systems we studied. In particular for $\Delta b_4$ and $\Delta b_5$
the subspace contributions enter with similar magnitudes but opposing signs in all the cases we studied, indicating
that the final answers for those coefficients are the result of potentially delicate, coupling-dependent cancellations.

Finally, it should be pointed that out, although we have shown results for a variety of attractively interacting Fermi gases, 
our analytic results apply to repulsively interacting cases as well. We defer the analysis of the repulsive case to future work.
To facilitate the application of our results to those and other cases, we have made our analytic formulas available in a Python 
code as Supplemental Material~\cite{SupMat}, along with data tables for the extrapolated virial coefficients.


\acknowledgments
We would like to thank Tilman Enss, Jesper Levinsen, Vudtiwat Ngampruetikorn, and Meera Parish
for providing us with their data and for early comments on aspects of this work.
This material is based upon work supported by the National Science Foundation under Grant No.
PHY{1452635} (Computational Physics Program).



\end{document}